\documentclass[conference]{IEEEtran}
\usepackage{cite}

\ifCLASSINFOpdf
  \usepackage[pdftex]{graphicx}
  \DeclareGraphicsExtensions{.pdf,.jpeg,.png}
\else
\fi

\usepackage{array}
\usepackage{xurl}
\usepackage{xcolor}
\usepackage[normalem]{ulem}
\usepackage{longtable}
\usepackage{subcaption}
\usepackage{multirow}
\usepackage{colortbl}
\usepackage{tabularx}
\usepackage{svg}
\let\labelindent\relax
\usepackage{enumitem}

\newcommand{\michal}[1]{{\em\color{blue}{}}}
\newcommand{\ian}[1]{{\em\color{orange}{}}}
\newcommand{\dali}[1]{{\em\color{purple}{}}}
\newcommand{\conor}[1]{{\em\color{cyan}{}}}
\newcommand{\todo}[1]{{\em\color{red}{}}}
\newcommand{\addition}[1]{{#1}}

\begin{document}
%
\title{
An analysis of scam baiting calls: \\
Identifying and extracting scam stages and scripts. 
}

\makeatletter
\newcommand{\linebreakand}{%
  \end{@IEEEauthorhalign}
  \hfill\mbox{}\par
  \mbox{}\hfill\begin{@IEEEauthorhalign}
}
\makeatother

\author{\IEEEauthorblockN{Ian D. Wood}
\IEEEauthorblockA{Macquarie University\\
ian.wood@mq.edu.au}
\and
\IEEEauthorblockN{Michal Kepkowski}
\IEEEauthorblockA{Macquarie University\\
michal.kepkowski@students.mq.edu.au}
\and
\IEEEauthorblockN{Leron Zinatullin} 
\IEEEauthorblockA{Macquarie University}
\linebreakand
\IEEEauthorblockN{Travis Darnley} 
\IEEEauthorblockA{Macquarie University}
\and
\IEEEauthorblockN{Mohamed Ali Kaafar} 
\IEEEauthorblockA{Macquarie University, Australia\\
dali.kaafar@mq.edu.au}
}


%


\IEEEoverridecommandlockouts
\makeatletter\def\@IEEEpubidpullup{6.5\baselineskip}\makeatother
\IEEEpubid{\parbox{\columnwidth}{
    Network and Distributed System Security (NDSS) Symposium 2024\\
    26 February - 1 March 2024, San Diego, CA, USA\\
    ISBN 1-891562-93-2\\
    https://dx.doi.org/10.14722/ndss.2024.23xxx\\
    www.ndss-symposium.org
}
\hspace{\columnsep}\makebox[\columnwidth]{}}

\maketitle

\begin{abstract}
  Phone scams remain a difficult problem to tackle due to the combination of protocol limitations, legal enforcement challenges and advances in technology enabling attackers to hide their identities and reduce costs. Scammers use social engineering techniques to manipulate victims into revealing their personal details, purchasing online vouchers or transferring funds, causing significant financial losses.
  This paper aims to establish a methodology with which to semi-automatically analyze scam calls and infer information about scammers, their scams and their strategies at scale. 
  Obtaining data for the study of scam calls is challenging, as true scam victims do not in general record their conversations. Instead, we draw from the community of ``scam baiters'' on YouTube: individuals who interact knowingly with phone scammers and publicly publish their conversations. These can not be considered as true scam calls, however they do provide a valuable opportunity to study scammer scripts and techniques, as the scammers are unaware that they are not speaking to a true scam victim for the bulk of the call. 
  We applied topic and time series modeling alongside emotion recognition to scammer utterances and found clear evidence of scripted scam progressions that matched our expectations from close reading. We identified social engineering techniques associated with identified script stages including the apparent use of emotion as a social engineering tool.  
  Our analyses provide new insights into strategies used by scammers and presents an effective methodology to infer such at scale. This work serves as a first step in building a better understanding of phone scam techniques, forming the ground work for more effective detection and prevention mechanisms that draw on a deeper understanding of the phone scam phenomenon. 
\end{abstract}


%

\section{Introduction}
Phone scams, sometimes referred to as voice phishing or ‘vishing’, are a form of social engineering attack that leverages the telephone system. 
Scammers generate persuasive scenarios to convince victims to share personal information or pay at times substantial sums of money. 
These can e.g.,~impersonate authoratitive sources such as tax or law enforcement, offer seemingly free gifts or present a imagined threat such as a hacker accessing your bank details. 
The prevalence of phone scams has increased dramatically in recent years \cite{cnn_ftc_nodate,snider_robocalls_nodate} flagging an urgent need for effective methods to combat them. 
Reports show that people suffer significant loses due to scams, losing, for example, \$1.2 billion to impostor scams in the US in 2020 alone \cite{noauthor_consumer_2021}. 
Victims of successful scams often feel embarrassment, guilt and shame, which is believed to contribute to the under-reporting of fraud cases \cite{noauthor_financial_2015}, with some demographics appearing disproportionately affected \cite{noauthor_protecting_2020,burnes_prevalence_2017,ellin_scammers_2019,deliema_financial_2020,hanoch_scams_2021}.

The problem of phone scams remains difficult to solve due to existing telecommunication system limitations (e.g.~the existence of legacy phone networks in the global phone system that do not support modern security measures) and techniques and technology used by bad actors (e.g. VoIP and Caller ID spoofing) \cite{sahin_sok_2017,noauthor_why_2021}. 
Scammers also frequently operate outside of the victim’s jurisdiction making it difficult to address from a legal and law enforcement perspective.

Considering scammers' monetary gains and difficulties with securing the telephone network, scam calls will likely remain a popular choice for criminals. 
Maintaining 
up-to-date data about the ongoing scam campaigns, scammer behavior and the techniques they employ is challenging and yet crucial to design and maintain effective defensive approaches. Data acquisition needs to be at once agile, able to adapt to the ever-changing scam landscape, to provide deep insights into the techniques employed in current scams and to do all this at scale. 
\addition{One particular application of such insights, and the initial motivation of this study, is the creation of conversational AI bots that masquerade as susceptible scam victims. Recent advances in conversational AI allow for fluent generation of language, however they still struggle with situational awareness, hence incorporation of live contextual knowledge in to such models promises to improve their ability to act as convincing scam victims. }
In this study we introduce a semi-automated framework to analyze scam calls and infer information about scammers,
their scam approaches and strategies at scale. 
We demonstrate our framework on a sample of recordings obtained from public sources, identifying insightful details on how scammers operate. 

To obtain scam transcripts, we searched YouTube for videos mentioning "scam call" or "scam call recording" as well as trawling the 
`scambaiting' YouTube tag. We transcribed the audio of the resulting videos using a commercial automated transcription service and further cleaned the data to remove irrelevant transcripts and sections of the transcripts that are not part of the scam call, leaving us with 341 transcripts totaling 90 hours of scam calls to analyze. 
Approximately 60\% of the 825 originally collected videos either did not contain actual conversations with scammers or were deemed of insufficient quality (e.g.:~containing only short disconnected snippets of scam conversation). 
Though this data does not constitute true conversations between scammers and their victims and represents a small sample of primarily US scam types, it does allow us to analyze scammer methodologies, including the scripts they follow and particular social engineering techniques they apply. Scam baiters deliberately draw out the call and present challenging personas to the scammers, providing a rich view on how the scammers themselves behave in a diversity of situations. 
\addition{Further, the actions of scam baiters and responses of scammers are a perfect match for the purpose of training and informing AI ``victim bots'', our initial and primary motivator.} 
It must be pointed out that scam baiters are not true victims, and though they may attempt to pose as such, likely do not present many behaviors that a true victim may exhibit. 
Thus our study can provide insights into the scripts and techniques used by scammers and some scammer behaviors, but cannot be considered comprehensive. 

The framework begins with the identification of phone scam types present in available data. 
We manually label all transcripts, 
identifying four main scam types and 3 further scam types with a marginal presence. 
Identifying the type of scam being undertaken in a call can be useful for gaining a deeper understanding of the mechanisms and techniques used by scammers. 
We then demonstrate recognition of the type of scam given a relatively small sample of annotated scams and simulated scam type recognition in a live call setting by limiting the number of utterances available to the recognition model. We found that identification of the scam type of a call is effective with just one utterance (80-90\% accuracy) and highly accurate with 5 or 6 utterances (92-98\%). 

Next, we propose an analysis of the content of each type of scam. 
We apply a contextualized topic model~\cite{Bianchi_Terragni_Hovy_2021} and a publicly available emotion detection model to scammer utterances. We find that the topic model is able to identify scam themes observed during close reading of scam transcripts and note relatively subdued emotions on the part of the scammer. 

During close reading of the transcripts, we observed consistent sequences of scam stages, with the scammers appearing to follow pre-defined scripts. In order to verify this observation, and as a further tool to automate the analysis and understanding of scammer techniques and the mechanisms they leverage, we apply a Hidden Markov Model (HMM) to the outputs of the topic model. 
HMMs attempt to uncover simple state transition processes underlying complex sequences of observations. 
We found that the HMM models are able to capture the structure and flow of the scam calls, again with clear interpretations that match close reading. This is an indication that the scammers use a consistent scam structure, likely following a well defined script. We verified the model through manual annotation and found model outputs to match human interpretations of identified scam stages. 
We then construct a machine learning model to infer the state in a live call setting (i.e.,~from the utterances up to a given point in a transcript, infer the scam state at that point). We were able to correctly infer the states with reasonable accuracy, achieving an increase of 45\% accuracy over a random model for scam types with more than 50 available transcripts, and around 80\% accuracy when we accept predictions that are one step ahead or behind. 
An overview of our processing and analysis pipeline can be found in Figure~\ref{fig:processing-pipeline}. 

The story thus revealed is scripted scam progressions with embedded social engineering techniques. For example, in social security number scams, the scammer first establishes themselves in a position of authority, posing as a representative of the social security administration or similar authority, then describes an ongoing investigation into highly illegal events (drug smuggling, money laundering, allusions to murder, ...) associated with the victim's social security number, all the while stating that their aim is to prove the victim innocent. The victim is asked to pay a fee to expedite the resolution process and avoid being prosecuted. 

To the best of our knowledge, this is the first work to  apply automated machine learning techniques to scam call recordings and effectively extract meaningful insights into scripts used by scammers and their strategies and behaviors. 
Our analysis allows us to paint a rich picture of the characteristics of scam calls, leading to a deeper understanding of the phenomenon. This work is intended as a first step toward future mitigation strategies and tools such as real time early detection and public education, leading to protection of potential victims and reduction of the amount of money lost to scams. 


The key contributions of this paper are as follows:

\begin{enumerate}[itemindent=0.0in,itemsep=1pt, topsep=1pt]
    \item
        We collect and collate a data set of 341 conversations between scam baiters and scammers, annotated by scam type. We publicly released this data set at \url{http:to.be.released.on.acceptance}.
    \item
        We develop and demonstrate a semi-automated framework for identifying and tracking current phone scam scripts, scammer behavior and scam strategies at scale. 
        Our framework leverages hidden Markov models combined with topic modeling and automated emotion recognition for identifying the progression of scam scripts and facilitating the recognition of persuasion techniques used in telephony scams. We demonstrate effective predictive models in a simulated live call setting. We publicly release the code for our predictive models at \url{http:to.be.released.on.acceptance}.

    \item
        We provide a summary of insights into the rich tapestry of social engineering techniques identified in our data through a combination of our automated analyses and close reading of the scam conversations. 
    
    
        
\end{enumerate}






\begin{figure*}[th!]
    \centering
    \includegraphics[width=0.8\textwidth]{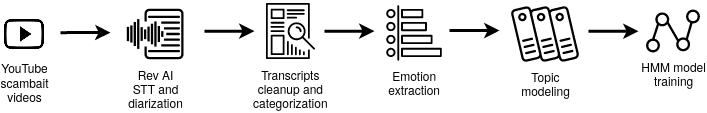}
    \caption{Data set sourcing and processing pipeline.}
    \label{fig:processing-pipeline}
\end{figure*}

\section{Related work}

\paragraph{Current solutions:}

Solutions to the problem of phone scams have been suggested for both telecommunications provider \cite{sheoran_nascent_2019,tu_toward_2016} and end user applications  \cite{reaves_authloop_2016,reaves_authenticall_2017,mustafa_end--end_2018} to combat Caller ID spoofing through authentication. However, as of the writing of this paper, these have not seen widespread adoption\addition{ in part because, to be effective, they would need to adopted universally across the globe}.

Existing deployed solutions focus primarily on creating blocklists of known bad numbers \cite{pandit_towards_2018} or proposing a reputation system based on caller behavior \cite{javed_detecting_2021,li_machine_2018,azad_socioscope_2020} yet many scam calls make it through despite these measures \cite{tu_sok_2016}. It has been shown, for example, that adding a number to the national Do-Not-Call Register may produce the opposite result \cite{sahin_effectiveness_2018} with attackers potentially abusing such lists.
Researchers have also proposed analyzing the audio features of the call \cite{balasubramaniyan_pindr0p_2010}, developing virtual assistants that vett incoming calls \cite{pandit_fighting_2020}, implementing application indicators that alert users to potentially unsafe calls \cite{sherman_are_2020} and using simple natural language processing techniques on initial call utterances to detect scams~\cite{derakhshan_detecting_2021}. Several commercial scam detection tools have recently appeared using these and related data driven techniques~\cite{noauthor_truecaller_nodate,noauthor_stop_nodate,noauthor_visual_nodate,noauthor_hiya_nodate,noauthor_about_nodate}. 

Despite these scam call prevention technology advancements, 2022 industry and government reports\cite{truecaller_report, commission_scam_2016, ukfinance_report} show that the scam calls as well as related monetary losses continue to increase (e.g., in 2022, over 30\% more money was lost than in 2021 as a result of the phone scams in Australia). 

\paragraph{Datasets and telephony honeypots:}

The ‘in the moment’ nature of telephone conversations makes it difficult to analyze them on a large scale. As a result, data used in research of scam calls to date has been limited in scale (e.g.~\cite{derakhshan_detecting_2021,sawa_detection_2016}) 
or containing primarily call metadata with no or little call content (e.g.,~\cite{li_machine_2018,javed_detecting_2021,prasad_whos_2020}). 

Telephony honeypots, systems deployed in telephony networks to capture malicious calls, are one option to scale up the study of malicious calls. 
Drawing on the use of honeypots to detect network intrusions, telephony honeypots have been deployed to detect voice spam \cite{gupta_phoneypot_2015,prasad_whos_2020,marzuoli_call_2016,prasad_diving_2023}. Similar methods were used to discover, record and analyze tech support scams \cite{miramirkhani_dial_2017}.
Despite some success of these methods, legal and regulatory constraints associated with recording calls in various jurisdictions as well as the need to engage with scammers in real time present challenges to this approach. 

\paragraph{Telemarketing and chatbots:}

There are some similarities between telemarketing and scam calls in respect to technology and delivery methods used, however attackers’ motivations, approaches and impact on victims differ when it comes to phone scams \cite{sahin_sok_2017}. 
Attempts have been made to waste telemarketers’ and scammers’ time using pre-recorded messages and chatbots \cite{sahin_using_2017,relieu_lenny_2019} however this approach has not been widely applied to countering scam calls.

\paragraph{Conversation analysis:}

Some preliminary analysis of phone scams has been performed \cite{maggi_are_2010,relieu_lenny_2019} highlighting the challenges with data collection compared to traditional phishing attempts. 
Researchers have studied persuasion techniques \cite{jones_how_2020} finding a similar diversity in persuasive techniques to our analysis with the exception that they identified \emph{social proof} as common where we did not find evidence of that technique. A further study used forensic linguistics \cite{tabron_linguistic_2016} to analyze scam calls so as to better understand methods used by attackers. 
More recently, the use of ‘scam signatures’ has been proposed \cite{derakhshan_detecting_2021}, paving the way for early detection of scams based on semantic content of the conversation.
Natural language processing techniques have been suggested to improve detection of phishing and other social engineering attempts in emails~\cite{lansley_seader_2020,peng_detecting_2018}, and phone scams~\cite{sawa_detection_2016}. 
Approaches to date, however, are not scalable, requiring substantial human input and tested on very small data sets.

\section{Data Set}
Our data consists of annotated and marked up transcripts of conversations between phone scammers and YouTube scam baiters. 
This section describes how we obtained and processed the data, covering the first 5 steps in Figure~\ref{fig:processing-pipeline}. The last step, ``HMM model training'' is discussed later in Section~\ref{sec:HMM}.

\begin{table*}[h!]
\footnotesize
\centering
\caption{Statistics of the scam baiters' transcripts}
\label{tab:scam_baiters}
\resizebox{\textwidth}{!}{%
\begin{tabular}{l|r|r|r|r|r|r|r|c|c}
\textbf{Scam Baiter} & 
\textbf{\begin{tabular}[c]{@{}l@{}}Trans- \\ cripts\end{tabular}} & \textbf{\begin{tabular}[c]{@{}l@{}}Scammer \\ utterances\end{tabular}} & \textbf{\begin{tabular}[c]{@{}l@{}}Scam Baiter \\ utterances\end{tabular}} & \textbf{\begin{tabular}[c]{@{}l@{}}Scammer \\ words\end{tabular}} & \textbf{\begin{tabular}[c]{@{}l@{}}Scam Baiter \\ words\end{tabular}} & \textbf{\begin{tabular}[c]{@{}l@{}}Scammer \\ duration (s)\end{tabular}} & \textbf{\begin{tabular}[c]{@{}l@{}}Scam Baiter \\ duration (s)\end{tabular}} & \textbf{\begin{tabular}[c]{@{}l@{}}Scammer word\\ rate (w/s)\end{tabular}} & \textbf{\begin{tabular}[c]{@{}l@{}}Scam Baiter word \\ rate (w/s)\end{tabular}} \\ \hline

Boda Scambaits  &         129 &               9460 &                   9460 &        133941 &            126335 &                53913 &                    48115 &                    2.49 &                        2.68 \\ \hline
Scammer Jammer  &          70 &               7855 &                   7855 &         94768 &            110570 &                38064 &                    48516 &                    2.50 &                        2.32 \\ \hline
Scammer Payback &          63 &               6691 &                   6691 &         99335 &             74496 &                34638 &                    26499 &                    2.88 &                        2.78 \\ \hline
scambait tv     &          33 &               2431 &                   2431 &         44156 &             23124 &                18949 &                    11771 &                    2.33 &                        2.03 \\ \hline
IRLrosie        &          18 &                872 &                    872 &          6468 &             14406 &                 3397 &                     6797 &                    1.95 &                        2.18 \\ \hline
Rinoa Poison    &          17 &               2858 &                   2858 &         36511 &             43402 &                12499 &                    14238 &                    2.86 &                        3.02 \\ \hline
Other           &          11 &                585 &                    585 &          7940 &              6864 &                 2757 &                     2507 &                    2.86 &                        2.81 \\ \hline
Total           &         341 &              30752 &                  30752 &        423119 &            399197 &               164217 &                   158443 &                         &                             \\

\end{tabular}}
\end{table*}

\subsection{Sourcing Scam Transcripts}


Collecting data for the purpose of studying scam calls is a challenging task as individuals do not typically record their phone calls and victims of scams may be hesitant to share or publicly release recordings of the calls due to potential embarrassment or concerns about exposing financial or personal information that were discussed during the call.

To obtain data for our analysis, we relied on recordings of phone scam conversations posted on YouTube by individuals known as "scam baiters". 
Scam baiters are individuals who engage with and record scam calls, attempting to draw out the call then typically confront the scammers about their unethical practices. 
While these calls cannot be considered representative of genuine interactions between scammers and their intended victims, the scammer utterances however are bona fide and provide insights into scammer techniques as well as the progression of hypothesised  scam call scripts. 



We searched YouTube for channels that mention ``scam baiting'' (or "scam baiters", "scammers baited", etc.) and manually vetted all their uploaded videos, carefully selecting those that predominately contained actual scam calls and with long, coherent scammer conversations. We filtered out instances of videos where scam baiters used irony or offensive language towards the scammers, and only kept the videos where the scam baiters seriously acted out the role of a scam victim. 


In this way, we obtained 341 transcripts primarily from six different scam baiting channels with an average length of 90.7 utterances (44.8 for scammers, 45.9 for scam baiters) and a median length of 82 (40.5 for scammers, 41 for scam baiters), 
and a maximum length 242 utterances. 



\subsection{Scam Baiters}
Our data set was collected from the publicly available recordings of conversations between scammers and scam baiters. Scam baiters pretend to be a vulnerable victim to engage scammers. Even though this does not reflect real scam calls, the analysis of how scam baiters interact with scammers can contribute to our understanding of the properties of scam calls.

In Figure~\ref{fig:scam-types} we see the distribution of scam types in our data. 
Interestingly, the most common scam types (reward, support, refund and social security number) were present in the recordings of all scam baiters. However, we can observe that scam baiters have a preference for a certain types of scam calls (e.g., Boda Scambaits published a significant number of scams about social security numbers).

Table~\ref{tab:scam_baiters} presents a detailed overview of the transcripts collected for each scam baiter.  Looking at the utterances lengths in words and seconds, we observed that 3 scam baiters (Scammer Jammer, Rinoa Poison and IRLrosie) try to overtalk the scammer, whereas the remaining 3 scam baiters allow the scammer to dominate the conversation.



\begin{figure}
    \centering
    \includegraphics[width=.48\textwidth]{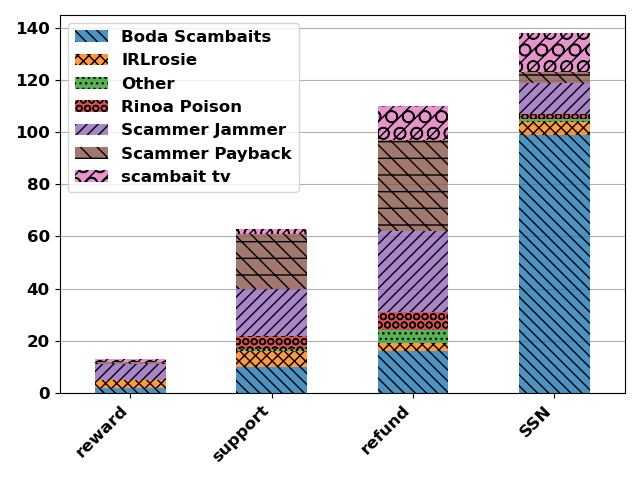}
    \caption{Scam types in collected scam baiting calls.}
    \label{fig:scam-types}
\end{figure}

\subsection{Data Pre-Processing}
We transcribed the YouTube videos into diarised text using a commercial speech-to-text API (rev.ai). 
We then manually edited the transcripts to assign utterances as either Scammer or Victim and to remove parts of the transcripts that could not be considered part of a typical scam call: YouTube video introduction, side comments by the host not relevant to the call itself, and the part of the conversation including and following the Victim reveal. 
Some transcripts were completely removed as, on closer inspection, they did not resemble a realistic scam call and instead were likely produced entirely for entertainment purposes. 

Due to the artificial nature of the "victim" in these calls, our analysis focuses on scammer utterances. Our final sample contains 15234 of these. 



\begin{table*}[ht!]
\footnotesize
\centering
\caption{Statistics of the collected Youtube scambaiting recordings transcripts }
\label{tbl:data-statistics}
\resizebox{\textwidth}{!}{%
\setlength\tabcolsep{2pt}%
\begin{tabularx}{1.05\textwidth}{l|r|r|r|r|r|r|r|c|c}

\multicolumn{1}{X}{\footnotesize\textbf{Scam type}} & 
\multicolumn{1}{|X}{\footnotesize\textbf{\# of calls}} & 
\multicolumn{1}{|X}{\footnotesize\textbf{Total \newline duration (s)}} & 
\multicolumn{1}{|X}{\footnotesize\textbf{Scammer  duration (s)}} & 
\multicolumn{1}{|X}{\footnotesize\textbf{Scam Baiter \newline duration (s)}} & 
\multicolumn{1}{|X}{\footnotesize\textbf{Total  words}} & 
\multicolumn{1}{|X}{\footnotesize\textbf{Scammer words}} & 
\multicolumn{1}{|X}{\footnotesize\textbf{Scam Baiter words}} & 
\multicolumn{1}{|X}{\footnotesize\textbf{Scammer words/second}} & 
\multicolumn{1}{|X}{\footnotesize\textbf{Scam Baiter  words/second}} \\ \hline

social security number &        138 &             119605 &                65446 &                    54159 &      309360 &        168935 &            140425 &                    2.55 &                        2.59 \\ \hline
refund                 &        110 &             115721 &                59744 &                    55967 &      295474 &        155167 &            140307 &                    2.61 &                        2.57 \\ \hline
support                &         63 &              61188 &                28180 &                    33003 &      153186 &         70392 &             82794 &                    2.42 &                        2.54 \\ \hline
reward                 &         13 &              12306 &                 4647 &                     7659 &       28326 &         11461 &             16865 &                    2.48 &                        2.33 \\ \hline
gift card              &         12 &              10599 &                 4879 &                     5720 &       26247 &         13201 &             13046 &                    2.47 &                        2.25 \\ \hline
family member          &          2 &               2282 &                  777 &                     1505 &        6553 &          2044 &              4509 &                    2.59 &                        3.02 \\ \hline
tax                    &          2 &                452 &                  274 &                      178 &        1556 &          1048 &               508 &                    3.82 &                        2.85 \\ \hline
charity                &          1 &                522 &                  270 &                      252 &        1614 &           871 &               743 &                    3.23 &                        2.95 \\ \hline
Total                  &        341 &             322675 &               164217 &                   158443 &      822316 &        423119 &            399197 &                         &                             \\ \hline
Total (hours)          &            &                179 &                   91 &                       88 &             &               &                   &                         &                             \\

\end{tabularx}}
\end{table*}

\section{Overview of the Data}

\subsection{Overall Description}
Here we present an overall description of the content of the captured scam transcripts informed by close reading and interpretation of data enrichment results (see Section~\ref{sec:data-enrichment}). 

In the sample analysed, scammers often mention or impersonate authority figures: IRS, HMRC, FBI, etc. Scammers also attempt to increase trust by providing a verification of stated claims or prove authenticity of the caller by using spoofed numbers that appear to originate from a known entity. 

In social security number scams, police, court orders, arrest warrants, gaol and other negative consequences in the event of non-compliance are mentioned. Scammers attempt to prevent interruption of the script or defer questions by talking over the victim. The victim is nudged towards a faster and easier option to avoid negative consequences by ‘resolving the matter’ here and now, introducing time pressure and preventing the victim from taking the time to think this through. 
The ``quick resolution'' is claimed to be available only if acted on immediately, introducing further time pressure. 
A decision is demanded during the call; a refund, reward or reversal is promised if the victim complies. 

Depending on the scam type, an attempt to establish the victim's location or request to travel to a location where the fee can be paid, gift card purchased or money deposited is often carried out. This is followed by a payment or purchase request or obtaining the victim's bank card details. Specific, not rounded up amounts are requested and legitimate sounding ways to pay are mentioned. 

Alternatively, rapport building and polite social protocol is followed in other scam types (e.g. tech support). Legitimate sounding explanations of the reason for the request and specific details are shared to add credibility to claims. 
Step-by-step guidance to navigate to a website, install software (often TeamViewer to get remote access to the victim's machine) or fill out online forms with personal and often financial information is provided.

A handover to another individual, often a supervisor, is sometimes introduced by scammers when a given milestone in the script is achieved (e.g. victim installs TeamViewer or confirms banking details).

Due to the length of some calls or if the desired result is not achieved immediately, scammers may attempt to end the current call and continue the scam in a subsequent call at a later stage.

\subsection{Data Statistics}
\label{sec:data-statistics}

Our data set contains 341 transcripts which in total represent 90 hours of phone scam conversations. We found 7 types of scam in the transcripts (see Section~\ref{sec:scam-type-annotation}), however, three of them, "family member", "tax" and "charity", have marginal presence, with only 2, 2 and 1 transcripts respectively and are not considered for further analysis. Social security number scam transcripts are the largest group in our dataset (140), followed by refund scams (110). The remaining transcripts belong to support (63) and reward (25) scams. Statistics on our data set can be found in Table \ref{tbl:data-statistics}. 

Firstly, we measured basic text statistics to learn what is the conversation approach of the parties (Scammer and Scam Baiter). Surprisingly, scammers do not always dominate the conversation. In terms of word count and duration, scammers used more words in social security number and refund scams, whereas Scam Baiters are more talkative in support and reward scam types. 
We can posit the presence of extended explanations on the part of scammers in these scams as the cause, however we acknowledge that active distraction from the Scammer's directions and story by Scam Baiters will also be a significant factor. 

Regarding word rate both Scammer and Scam Baiter are close to the regular speech rate range (2 – 2.5 wps\footnote{According to  National Center for Voice and Speech \url{https://ncvs.org/archive/research_tissue.html}}). 
There appears to be a level of word rate coordination: higher scam baiter word rates matched with higher scammer word rates. This is of interest, as it indicates a level of connection, where the scam baiter may be affecting the scammer, and warrants further investigation.


\section{Data Enrichment}\label{sec:data-enrichment}
In this section we describe the modeling and annotation we performed to enrich the data. This included manually annotating scam types for each transcript, emotion extraction per utterance and automated extraction of scam sequences, which can be thought of as  a proxy for scripts followed by scammers. Scam sequence extraction was done through a combination of topic modeling to extract themes and common word patterns followed by hidden Markov modeling (HMM) over topics to extract scripted and thematic sequences. 

\subsection{Scam Type Annotation} \label{sec:scam-type-annotation}
We annotated the overall type of scam for each transcript in our data. 
Two of the papers' authors conducted the annotation. First a classification scheme was independently determined, then a consensus view of the scam types present was taken --- the scam types mentioned in Section~\ref{sec:data-statistics} were unambiguous and easy to agree on, with the exception of ``reward'' scams, which were initially identified as two categories and later merged (see below). 
An initial 82 transcripts were then annotated independently by both authors with the agreed on scam types. Initial annotations had 83\% agreement, and all discrepancies were readily identified as simple errors or ambiguities in the interpretation of two of the categories. It was decided that these two categories (``gift card'' and ``reward'') should be merged as they represented minor variations on the same scam sequence and both had few transcripts. 
The remaining transcripts were annotated with the updated scheme by one author and verified by the second. 





\subsection{Emotion Extraction}

\begin{figure*}[ht!]
    \centering
    \includegraphics[width=0.8\textwidth]{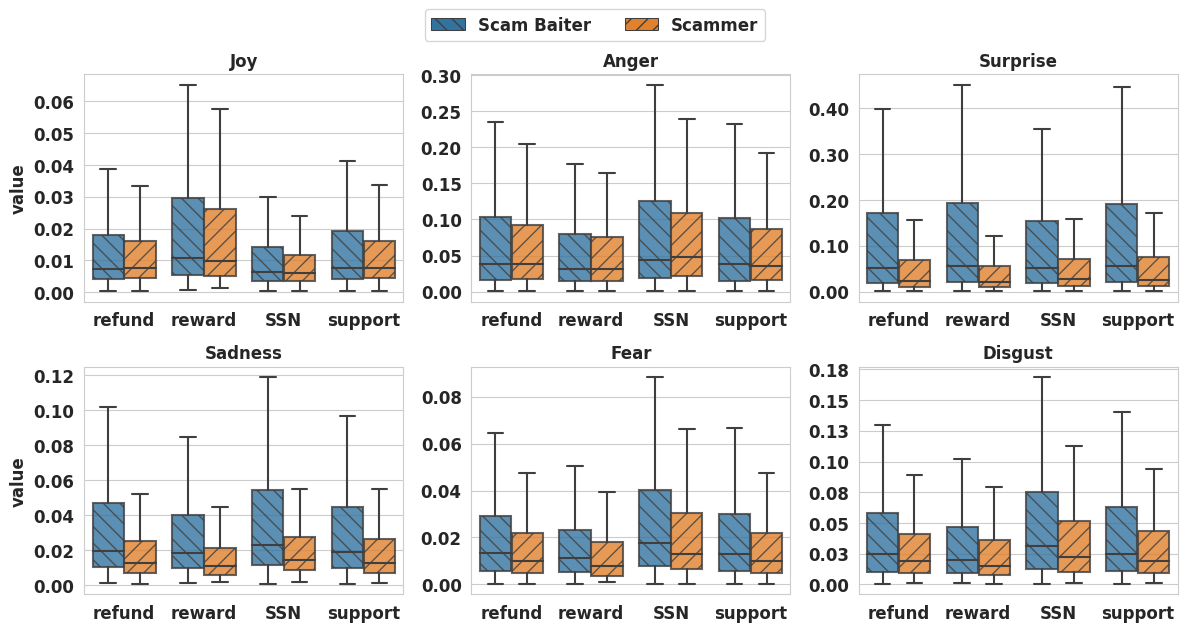}
    \caption{Ekman's 6 basic emotion scores for each scam type.}
    \label{fig:scam-types-emotions}
\end{figure*}

For emotion detection we used a RoBERTA~\cite{liu_roberta_2019} based emotion prediction model trained on a balanced subset of almost 20,000 human annotations from 6 publicly available data sets\footnote{https://huggingface.co/j-hartmann/emotion-english-distilroberta-base}. 
The model estimates Ekman's 6 basic emotions plus neutral: joy, sadness, anger, fear, disgust, surprise and neutral. 
The model was trained using cross-entropy loss on discrete labels (0 if an emotion is not present, 1 if it is). As such the scores can be interpreted as probabilities that the given emotion is present. 

Figure \ref{fig:scam-types-emotions} presents distributions of emotion scores by scam type. We found that for all scam types, surprise and anger are the most common emotion. 
We note that overall, the scammers were less emotional than the scam baiters. This is particularly visible with surprise. We propose two complementary explanations for this. First, the primary scam types present in our data present serious scenarios: identity theft (social security number scams), ongoing hacking of the victim's computer (support scams), and scammers posing as company representatives (refund and reward scams). In those roles, it is natural that the scammer would be serious and unemotional, but this also plays a role in social engineering, increasing the sense of authority. 
Second, on listening to the calls, it was our impression that that the Scam Baiters were acting out exaggerated roles, in part to entertain their audience and in part to engage the scammers. It is unclear how their portrayals would relate to the emotions of true scam victims and potential victims. 



\subsection{Scam Progression} \label{sec:extracting-scam-scripts}
In order to understand the methods and tools of phone scammers, a deeper understanding of how scam calls progress is needed. 
Here we seek to establish automated or semi-automated approaches to analyze and track scam call content, and with a view to deployment in a large scale scam call monitoring setting. 
We first seek a high level understanding of the content of scammer utterances using recent topic modeling methods. We find that the topics thus uncovered faithfully reflect our observations from close reading, identifying key steps that we observed in the transcripts. The topic model is applied to all the data with individual utterances as documents. 

Each transcript is then transformed into a sequence of topic intensities, and we apply a Hidden Markov Model (HMM) over these sequences for each scam type. Again, the states inferred by the HMM models correspond to our close reading observations, revealing clear scam progressions that match our observations. As can be expected, the quality and richness of HMM models varies with the quantity of data available. 
Finally, investigate the feasibility of identifying the scam stages as identified by the HMM models in a live call setting. 

\subsubsection{Topic Modelling}
\label{sec:topic-modelling}

We used contextualised topic models~\cite{Bianchi_Terragni_Hovy_2021}, a recent neural topic model combining contextualised representations from large pre-trained language models and the variational auto-encoding neural topic model  ProdLDA~\cite{Srivastava_Sutton_2016}. 
We trained a model with 50 topics on scammer utterances. The model showed good topic diversity (inverse rank biased overlap~\cite{Bianchi_Terragni_Hovy_2021} 0.993) and reasonable coherence ($C_V$ coherence~\cite{Roder_Both_Hinneburg_2015} of 0.454). We explored $\beta$ values of 0.01, 0.1, 1.0 and 3, and found 1.0 to perform well on both $C_V$ and $NPMI$ coherence metrics. 

We manually labelled each topic by considering both the top topic words and observing the top utterances (those scoring highest on the topic) and a weighted sample of utterances (using topic scores as weights) from the top 100 utterances. 
In most cases, consistent and readily interpretable semantics were observed, with a small number of topics appearing to merge distinct meanings (6 topics) and a small number with no apparent coherent meaning (2 topics). Table~\ref{tab:topic-freq-top} lists the most prominent topics for each scam type. See Appendix~\ref{appendix:topic-desc} for a full list of estimated topic labels and indications of merged/incoherent topics. 
Overall, we found that the topic model successfully revealed semantics that we observed in the data. 

Finally, we discuss topics and their frequency of appearance in utterances of each scam type. The top topics (with probability above 1.5x the average topic probability of  0.02) and their frequencies are shown in Table \ref{tab:topic-freq-top}. The complete topic frequency data can be found in Appendix  \ref{appendix:topic-desc}. For the social security scams, we found 6 topics with elevated probability. For example, topics 32 and 39 that address legal charges against the victim. Similarly, we found elevated probability for topics that describe vouchers and gift cards with reward scams (topics 6 and 38). In the case of support scams, the 2 topics with elevated probability (8 and 22) focus on technical details such as computer or phone instructions. Interestingly, refund scams do not have topics with elevated representation above our chosen threshold, and we show the top two topics. We note that the highest topic frequencies for support and refund scams are relatively low. We believe this is due to the lower quantity of data from these scam types and the observed diversity in the scripts and techniques used. 

\begin{table}[t]
    \centering
    \caption{Top topics and their frequencies in scams of each type. Average topic frequency is 0.02.}
    \label{tab:topic-freq-top}
    \renewcommand{\arraystretch}{1.2}
    \resizebox{\columnwidth}{!}{%
    \begin{tabular}{l|c|c|p{160pt}}
         \textbf{Scam type}& \textbf{Topic id} & \textbf{Frequency} & \textbf{Topic short description} \\
         \hline
         \multirow{4}{*}{Reward} 
         & 38 & 0.063 & small payment/fee\\
         \cline{2-4}
       & 6 & 0.054 & rebate/voucher\\
       \cline{2-4}
       & 3 & 0.047 & obtain gift card \\
       \cline{2-4}
       & 4 & 0.039 & deals and discount vouchers\\
       \hline
       \multirow{5}{1.5cm}{Social security number} & 43 & 0.034 & car in Texas, blood, drugs \\ 
       \cline{2-4}
    & 44 & 0.034 & identity theft questions\\
    \cline{2-4}
    & 39 & 0.033 & legal charges, call monitored, ``don't interrupt''\\
    \cline{2-4}
    & 32 & 0.033 & legal enforcement action \\
    \cline{2-4}
    & 25 & 0.031 & request/provide case id, verify DOB/SSN\\
    \hline
    \multirow{2}{*}{Support} 
    & 22 & 0.031 & hacked, secure your computer \\ 
       \cline{2-4}
    & 8 & 0.030 & payment instructions \\
    \hline
    \multirow{2}{*}{Refund} 

    & 16 & 0.029 & click instructions, free download \\ 
       \cline{2-4}
    & 8 & 0.029 & computer instructions (team viewer) \\
     \hline
       
    \end{tabular}}
\end{table}

\subsubsection{Hidden Markov Models over Topics}
\label{sec:HMM}
\begin{figure*}[ht!]
    \centering
    \includegraphics[width=\linewidth]{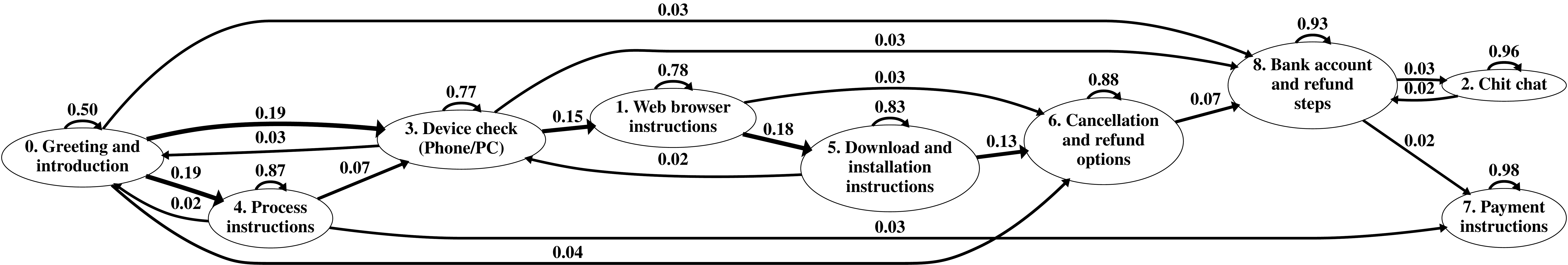}
    \caption{HMM state transition graph for ``refund'' scams. Numbers on links indicate transition probabilities (those less than 0.021 omitted\protect\footnotemark).}
    \label{fig:refund-transition-graph}
\end{figure*}

\addition{We use a hidden Markov model to extract scam scripts presumed to be followed by the scammers for each scam type. 
Hidden Markov models seek to represent time sequence data as the trace of a finite state machine. States ``generate'' utterances through a list of topic probabilities (known as emission probabilities) associated with each state and there is a table of state transition probabilities which determine the state of the subsequent utterance. HMM inference algorithms attempt to find a set of states (given by emission probabilities) and a state transition table which maximises the probability that the data was generated by those states and transitions. Typically the number of states is fixed. }

We use the multinomial hidden Markov model from the python package \emph{hmmlearn} with default parameters. The data consisted of the top topic (from our 50 topic model) for each utterance, providing a sequence of integers (one for each utterance) for each transcript. 

First, we split the data into 3 equal subsets (train, validation and test) and and made an initial selection of the number of states for each scam type through grid search, ranking by log likelihood on validation data. We observed that the log likelihood scores on test data with the resulting models aligned with those on validation data and were marginally better than on training data, indicating that the model is not overfitting. 
We then used 5-way cross validation on all data to make a final choice of the best number of states for each scam type\footnote{For support and reward scams we took the second best (7 vs. 4 states and 5 vs. 3 states respectively) as the log likelihoods were similar and resulting transition graphs more informative.}, then trained our final models using all available data. In all cases, we fit each model configuration 50 times, taking the model with best log likelihood (on validation data for cross validation and on all data for the final models). This is usual practice with this kind of HMM model, as the gradient descent based EM (Expectation - Maximization) algorithm has a tendency to get stuck in local minima --- multiple random starts usually results in a much better fit. 

To evaluate the models, we first examined the apparent meaning and coherence of inferred states. 
We manually labeled each inferred state by considering the labels and utterances associated with the most significant topics linked to each state. 
Overall, we found that a consistent interpretation could be applied to the utterances associated to a given state and that the progression of states is readily interpretable as a scam script progression and agrees with our observations from close reading. 
An example social security number transcript with utterance states is provided in Appendix~\ref{appendix:hmm-transcript-example}. 
Figure~\ref{fig:refund-transition-graph} shows the resulting graph for refund scams, see Appendix~\ref{appendix:transition-graphs} for the remainder. 

We then chose one transcript from each scam type at random and three authors manually annotated utterances using our state labels and descriptions of associated topics as a guide. We allowed two state choices where there was some ambiguity and considered annotations to agree if the second choice matched the first choice of other annotators. 
Table~\ref{tab:krippendorf} shows Krippendorff's Alpha between annotators and Cohen's Kappa between states inferred by HMM models and those obtained by vote between annotators, alongside the number of utterances in the selected transcripts and number of states used in both annotation and HMM models. 
We found that annotators could consistently identify states, with strong inter-annotator agreement for Refund and Reward and moderate agreement for social security number and support scams. Kappa scores show substantial agreement with HMM inferred states for all but the Reward scam, which has moderate agreement. 

\begin{table}[ht!]
\footnotesize
    \centering
    \caption{Annotator reliability (Krippendorff's Alpha) and agreement with HMM (Cohen's Kappa) and number of utterances (Utt.) for selected transcripts.}
    \label{tab:krippendorf}
    \begin{tabular}{c|c|c|c|c|c}
               &            &        &  & \multicolumn{2}{c}{\textbf{Kappa vs. HMM}}\\ 
    \textbf{Scam type}  &     \textbf{Utt.}   & \textbf{States} & \textbf{Alpha}           & \textbf{(strict)} & \textbf{(relaxed)} \\ \hline 
       SSN     &      33    &   11   &  0.57           &    0.55    &    0.66     \\
       Refund  &      66    &    9   &  0.80           &    0.52    &    0.65     \\
       Support &      68    &    7   &  0.51           &    0.61    &    0.69     \\
       Reward  &      21    &    5   &  0.70           &    0.28    &    0.50     \\
    \end{tabular}
\end{table}


Scam types with more data provided HMM models with more states that were also more expressive and easier to interpret. This reflects the ability of statistical based dimensionality reduction models such as HMMs and topic models to reliably extract structures in the data. With more data, there is sufficient evidence for more nuanced models without overfitting. 
Another feature of the graphs to note is the generally lower probabilities assigned to transitions in the later stages. This is particularly prominent with states 3 and 4 with social security number scams (see Appendix~\ref{appendix:hmm-state-interpretations}), which appeared mostly at the end of transcripts and whose links did not pass the threshold for inclusion in the graphs. We believe this is due to two factors: first, the fact that a significant proportion of the transcripts end before the later stages of the presumed scam script, hence there is relatively little data, resulting in less focused states (covering broader semantics) with fewer transitions. The second factor is the greater diversity in conversations in later stages of the scam, where the scammer adapts their instructions to the victims circumstances and the scam baiter has had more opportunity to derail the script. We observed numerous transcripts where the conversation diverged into random and extended chit-chat --- essentially, the scam baiter had succeeded in distracting the scammer from the script. 

Another important methodological point is the value of building these HMM models on data from a single type of scam, which substantially improves the clarity of the resulting model. We at first built a model on all our data, and although it was not without interesting and interpretable structures, the models we present here are much more focused and more clearly interpretable. This highlights the importance of classifying and labeling scam types. We note that topic models are more robust to diverse semantics (indeed are designed to tease them out), and being unsupervised, generally require more data to obtain a good model, thus we chose to pool data across scam types for the topic model. 

The coherence of the structures in the underlying data also plays a role here, and we take the success of the HMM approach, together with the sequential nature of the resulting models as indicators that there are well defined sequential structures consistent across the data, supporting the hypothesis that the scammers are following pre-defined scam scripts. 
We further discuss interpretation of the inferred HMM models from the perspective of social engineering in Section~\ref{sec:social-engineering}.

\section{Observations of Social Engineering}
\label{sec:social-engineering}
The term ``social engineering'' covers persuasive techniques used by scammers to convince their victims to act to their own detriment and to the scammers benefit. 
In this section we discuss social engineering techniques observed through close reading of our data and topics and HMM states associated with those techniques. 

The art of persuasion has been studied over many years, and several categorisations of psychological principles that can enable a manipulator to control the actions and choices of victims have emerged. In the words of pioneering psychologist in this area, B. J. Fogg ``As I see it, social influence is a broad area, with flexible boundaries and competing ways to categorise influence strategies''~\cite{Fogg2008Mass}. 
Keeping that in mind and for the purposes of simplicity, we organise our discussion of social engineering techniques following the five categories provided by Ferreira et.al.~\cite{Ferreira2015Principles}, which draws together two previous categorisations~\cite{Cialdini2006Influence,Weiksner2008Six} into a single model and has a focus on techniques used in phishing, taking careful note of boundary cases and those that do not really fit.


We investigated the data through a combination of close reading and interpretation of topic model topics and HMM states seeking to identify the application of social engineering by the scammers. 
Overall, we found that the scam calls are \emph{very} rich in applied social engineering techniques, with a substantial proportion of scammer utterances building on the persuasive landscape of the scams. It is not the purpose of this work to provide a comprehensive review of the techniques used in our data, however we provide here examples of the more prominent uses of persuasive techniques we encountered, and where practical, link them to topics and HMM states. 

We also note that in many cases, multiple persuasive techniques are combined in a single scam step or even in a single utterance. 
This is not unusual in persuasive acts, for example the invitation messages of successful Facebook apps often achieve this~\cite{Fogg2008Mass}. 
One example is this quote from a social security number scam: \textit{\ldots be specific and genuine on this phone call because this recording can be used in your favor or can be used against you in the courthouse\ldots}. Here we see Authority (referring to the court house), distraction (fear of legal consequences) and commitment (once the victim provides the requested PII).

\begin{quote}
    \emph{Authority (AUTH):} 
    Society trains people not to question authority so they are conditioned to respond to it. People usually follow an expert or pretence of authority and do a great deal for someone they think is an authority.~\cite{Ferreira2015Principles}
\end{quote}
This is a common persuasive technique used in scams as has previously been observed (e.g.~\cite{jones_how_2020}), and our data is no exception. We observed the scammers assuming the role of an officer from the social security administration in social security scams (\textit{\ldots you for calling social security administration\ldots}, \textit{\ldots My name is officer Carol Snyder\ldots}), or a qualified IT support professional in support scams, of an employee from a reputable company in reward scams (\textit{\ldots Thank you for calling PayPal. This is Daniel. How can I help you?\ldots}). 

Further subtle markers of authority are also present in several scam types. Examples include the scammer ``verifying the identity'' of the victim (e.g. state 7 and topic 46, \textit{\ldots Can you verify me the last four digits of social security number?\ldots}) as well as formal sounding language and procedures (\textit{\ldots This is the case identification number\ldots}). 

\begin{quote}
    \emph{Social Poof (SP):} 
    People tend to mimic what the majority of people do or seem to be doing. People let their guard and suspicion down when everyone else appears to share the same behaviours and risks. In this way, they will not be held solely responsible for their actions.~\cite{Ferreira2015Principles}
\end{quote}
In the scam types present in our data, this technique did not appear to play a significant role. 

\begin{quote}
    \emph{Liking, Similarity \& Deception (LSD):} 
    People prefer to abide to whom (they think) they know or like, or to whom they are similar to or familiar with, as well as attracted to.~\cite{Ferreira2015Principles} 
\end{quote}
In all scam types, we find HMM states that resemble informal chit-chat. The willingness of scammers to engage in friendly conversation with their victims appears to be an application of this technique. This is suggested in particular by the common proximity of chit-chat with payment instructions and procedures, a critical point in a scam at which the trust of victim is crucial.  

In some transcripts, the scammer and scam baiter already know each other from previous calls, meaning the scam spans several days and multiple calls. This longer engagement engenders familiarity and trust and represents the application of this technique. 

\begin{quote}
    \emph{Commitment, Reciprocation \& Consistency (CRC):} 
    People feel more confident in their decision once they commit (publicly) to a specific action and need to follow it through until the end. This is true whether in the workplace, or in a situation when their action is illegal. People have tendency to believe what others say and need, and they want to appear consistent in what they do, for instance, when they owe a favour. There is an automatic response of repaying a favour.~\cite{Ferreira2015Principles} 
\end{quote}
This category is rather broad, and covers multiple categories from other categorisation schemes. 

Reciprocation: Scammers will often state that they are attempting to help the victim (\textit{\ldots So let me help you to get connected with our Amazon\ldots} --- refund scam; \textit{\ldots I just wanting to help you out to get there.\ldots} --- support scam; \textit{\ldots There is no need to worry because I'm here to help you\ldots} --- social security number), which represents an appeal to Reciprocity. 

Commitment: In all the main scam types in our data, after explaining the situation, the scammer asks if the victim would like to fix the problem (social security number, support and refund)  or receive the gift (reward). By answering ``yes'', the victim is subsequently motivated to proceed with the scam rather than go back on their commitment. 

An involved sequence of tasks for the victim is also a form of commitment, where the victim has invested in and implicitly validated the scenario presented by the scammer. 
In social security number scams, the victim is asked to verify their identity (state 7), write down the case id (state 10), listen to the litany of evidence against them (states 1, 0). 
In refund scams, victims are asked to provide information about their phone/computer (state 3) and download and install remote desktop software (state 5).

One question of note that appears in many social security scams asks the victim for approximate bank balances (\textit{\ldots what will be the current dollar amount balance approximately you having in your checking account and as well as in your savings?\ldots}). We believe this also serves to inform the scammer of the value of the victim --- particularly high value victims are afforded more time and effort, and are often redirected to more senior scammers.

\begin{quote}
    \emph{Distraction (DIS):} People focus on one thing and ignore other things that may happen without them noticing; they focus attention on what they can gain, what they need, what they can lose or miss out on, or if that thing will soon be unavailable, has been censored, restricted or will be more expensive later. These distractions can heighten people’s emotional state and make them forget other logical facts to consider when making decisions.~\cite{Ferreira2015Principles}
\end{quote}
All scam types in our data utilise a form of distraction --- a (typically highly emotional, typically fearful) scenario that captures the victims attention such that they focus less on the steps they are asked to take.

In support scams, the scammer reports that there has been suspicious activity on the victim's computer and goes on to claim that it has been hacked and the hackers can steal their bank details etc\ldots 

In refund scams, they claim that a victim's online purchasing account has had an expensive order that was suspicious (\textit{\ldots This is suspicious activity on the Amazon account\ldots}).

In social security number scams, the whole scam scenario of atrocious acts combining drug smuggling, money laundering (\textit{\ldots these are now used to launder approximately \$240,000\ldots}) and hints of murder, all carried out using the victim's stolen identity can undoubtedly cause extreme fear and apprehension. 

In reward scams, a free gift is offered. 



\section{Other Insights into Scammer Methods}
\subsection{Scam Stages}
Our HMM models painted an interesting picture of scam call progressions. All analysed scam types have a similar high level structure: 
\begin{enumerate}[noitemsep, topsep=1pt]
    \item Greetings.
    \item Explanations of the problem.
    \item Instructions on what to do.
    \item Financial exploitation attempt.
\end{enumerate}
The exception is the HHM model for reward scams which was not as successful at identifying scam structure, likely due to lack of data (see Section~\ref{sec:HMM}). 
For example, in refund scams (Figure \ref{fig:refund-transition-graph}) greeting is represented by state 0., problem explanation is covered by state 4. States 3., 1., and 5. describe instructions on what to do and states 6., 8., and 7. target financial exploitation.  
Note that the greetings for each scam type are distinct, with the scammers introducing themselves as different characters, using different levels of formality etc\ldots, which enabled distinct scam types to be distinguished on the basis of very few utterances (see Section~\ref{sec:predict-scam-type}). 

Interestingly, we noticed that in 3 models (social security number, refund, and support) the state representing casual chit chat is linked with the money/payment related state. The explanation for this can be twofold. We suspect that it may be the scammer's deliberate action to distract the victim from the actual target of the scam (i.e., stealing money --- see Section~\ref{sec:social-engineering}). The second option is that scam baiters actively change the topic of the conversation when the scammer tries to close the scamming attempt with a financial request. It is likely that both have an element of truth. 

We noticed that one of the states for social security number scams adds an interesting variation of the scam that is not present in other types. The "redirection to supervisor" state (state 6.) is used to redirect the victim to a second scammer. We suspect that this technique is effective in the social security administration context because of the authoritative position played by the second scammer (usually introduced as senior agent). It would also allow senior, more experienced scammers to be brought to the call for the final stages that lead to payment.


\subsection{Weaponized Emotions}
Phone scams, unlike other types such as email or SMS, the scammer has greater presence and thus a more direct opportunity to impact victim emotions. 
From our preliminary examination of data set recordings, we observed that scammers leverage emotion manipulation in their social engineering techniques. Here we present a review of emotions detected in our data as seen from the perspective of inferred states and topics, and attempt to identify the roles they played in the scams, if any. 
Though we observed that scammers expressed less emotion than scam baiters, none the less variations in how scammers express emotion were evident. 

\begin{figure}[t]
    \centering
    \includegraphics[width=.48\textwidth]{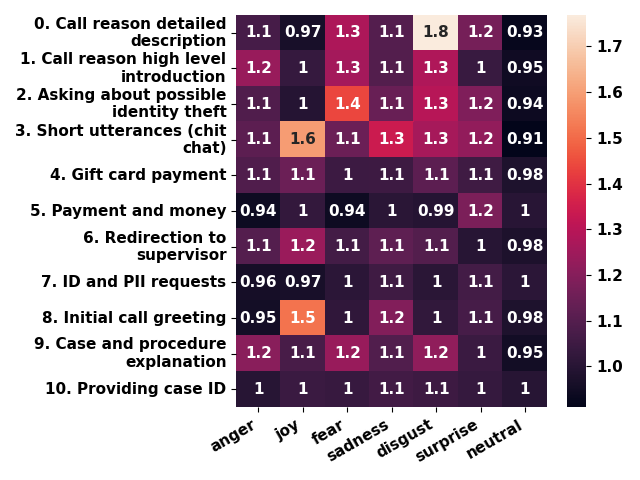}

    \caption{Social security number scam emotion heat map for HMM states. Scores are mean for that state relative to median over all utterances. }
            \label{fig:ssn-emo-heatmap}
\end{figure}

To measure which state is associated with a particular emotion, we averaged the emotion scores for utterances assigned to the state by the HMM model. 
Figure~\ref{fig:ssn-emo-heatmap} and Appendix~\ref{appendix:emo-state-heat-maps} present heat maps of relative emotion strengths for each state (relative to the median emotion strength among all utterances for that emotion), providing an indication of where scammer emotion is concentrated. 
We observe that in many cases there is a concentration of emotion on particular HMM states. 
This suggests that scammers may indeed use emotions in their social engineering techniques. 

In order to interpret the concentrations of emotion for social security number scams, we examine the highest emotion scoring utterances for each state-emotion pair with higher relative emotion score (see Figure~\ref{fig:ssn-emo-heatmap}). 
Firstly, we examined anger, which was high in "Call reason high level introduction" (state 1.) and "Case and procedure explanation" (state 9.). We found that scammers try to limit the questions from the victim by making sure they are not interrupted. We found numerous of examples of phrases that angrily enforce victims to not interrupt (e.g., \textit{\ldots Do not interrupt me in between\ldots, \dots Listen to me, do not interrupt me once more\ldots}, \textit{\ldots suspend your social for interrupting the officer\dots}). 
We also noticed that angry language was used in threats presumably to encourage compliance. For example, \textit{\ldots I will send a local sheriff on your doorstep\ldots}, \textit{\ldots I'm sending the cops to your house\ldots}. 

The score for the joy emotion is notably high for "Short utterances (chit chat)" (state 3.) and "Initial call greeting" (state 8.). In the case of initial call greetings, scammers tend to initiate the call with a joyful phrase (e.g., \textit{\ldots Hello. And thank you for calling\ldots}). 
For the chit chat utterances (state 3) we could not find any common scheme, though we note that chit chat utterances were more emotional overall. This is consistent with the scammer leveraging the victim's desire for connection to increase rapport (the LSD category of social engineering techniques --- see Section~\ref{sec:social-engineering}). 

The results for fear clearly show that utterances used to explain the scam theme are meant to trigger fear (states 0,1,2,9). The highest score was found for state 2 - "Asking about possible identity theft" (e.g., \textit{Have you ever lost it or someone stolen your personal identities from you in Texas? Like your driving license, your social security card, or any of your state ID?}). We found scammers suggest that the identity was used in illegal operations, most likely to increase seriousness of the case  (e.g., \textit{\ldots your social security number \ldots has been found suspicious for criminal activity."}). 

Sadness was the most prominent in chit chat utterances (state 3.). We suspect this is because of more compassionate language including apology phrases (e.g., \textit{\ldots I'm very sorry for the officers who didn't explain to you the case\ldots, \textit{I'm really sorry for the inconvenience, sir}\ldots, \textit{I'm really sorry. I cannot assure you}\ldots}), though empathizing with victim sadness also plays a role, for example [after Victim talks about passed husband and family]\ldots \textit{They are so lucky to have you, so yeah. And you'll live by yourself}\ldots

An outstandingly high score (1.8) was found for disgust  for the ``Call reason detailed description" state. The utterances for this state and emotion contain descriptions of crime scenes such as stolen cars, drugs and blood (e.g., \textit{\ldots the investigation started when we found an abandoned car \ldots and the car contained some blood as well as some drugs \ldots}). Interestingly, surprise scores are relatively consistent between states, with no states standing out, and neutral scores are especially so. 

\section{Applications: Predicting Scam Type and Scam Stages}\label{sec:predict-scam-type}
\addition{Incorporating side information into text based NLP systems is well established. In our setting, conversational AI agents can be tailored to specific types of scam, leveraging extracted scam scripts for known scam types. In order to make best use of this knowledge, we need to detect which type of scam we are seeing early in a call and we need to be able to track how and when the call progresses through the script. }

\subsection{Detecting Scam Type}
We wish to predict the type of scam, preferably early in a call, in a simulated live call setting. This is achieved by progressively restricting the number of utterances available to the predictive model. Note that we do not attempt to distinguish legitimate calls from scam calls in this work. 
We use a standard text classifier built on a RoBERTa base model~\cite{liu_roberta_2019} from the Huggingface model repository\footnote{\url{https://huggingface.co/roberta-base}} and build a separate binary model for each scam type. We use only scammer utterances for this task. 

\begin{figure}[!ht]
    \centering
    \includegraphics[width=\columnwidth]{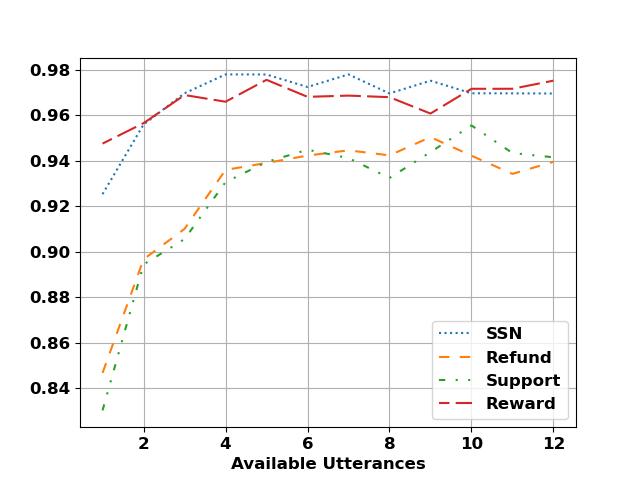}
    \caption{Scam type prediction performance (F1). }
    \label{fig:type-prediction-ssn}
\end{figure}

Figure~\ref{fig:type-prediction-ssn} shows recognition performance for the four scam types we investigate. We report F1 scores for the predictors; Precision, recall and accuracy follow similar trajectories. 
All scam types have strong performance with access to just one utterance (85\% - 95\%) and very strong performance with 5-6 utterances (92\% to 98\%). 
Refund and support scams are harder to distinguish than social security number and reward scams. 
We hypothesise that this is due to greater diversity in the conversations with scam baiters and in variations in specific details of the scams themselves. 

These results match our expectation and demonstrate that effective tools can be built for scam intelligence gathering and applications that operate on live scam calls based on identifying scam types early in a call. 
There are two main caveats that should be noted here. Firstly, our approach assumes that it is known that a given call is a scam call, however there are effective approaches and tools for detecting scam calls in a live call setting, so we argue that this is not a limitation. Secondly, only four scam types are included here, and the task we perform distinguishes between those four. There are, however, many and diverse phone scams being deployed in the world. We demonstrate that it is possible to distinguish between diverse scam types, as represented by our data, however as scams evolve and borrow from each other, we may expect that some scams (and hence, for example, the call centers they operate from) will be more difficult to distinguish.  

\subsection{Predicting Scam Stages}
Here again we deal with a simulated live call setting. \addition{We assume we already know which type of scam we are dealing with and wish to track the progression of the call through an already established scam script such as those extracted through the techniques in Section~\ref{sec:extracting-scam-scripts}.} 
To demonstrate the feasibility of the task, we wish to predict the stage of a scam call in a simulated live call setting. We train predictive models whose inputs are sequences of scammer utterances up to a point in the transcript and whose target is the scam stage for that utterance as inferred by the appropriate HMM model. We do this separately for each scam type. Since approximate identification of the current position in the scam script is also useful information, we consider a relaxed target that includes the previous scam stage and the succeeding one --- that is, the scam stage of the first utterance after (or preceding) the utterance in consideration that has a different scam stage. 
See Appendix~\ref{appendix:nlp-model-details} for details data preparation, model training and evaluation. 

Our models achieve a margin of 50\% over a random model for social security number and refund scams, and 30\% over for support and reward scams (Table~\ref{tbl:predict-scam-stage}). Interestingly, this margin is similar for the strict and relaxed targets. Again we see that support and reward, with less data available, have a lower margin over a random model. 
It is interesting to note that performance on the strict target is similar across scam types. This at fist seems perplexing, as the task for e.g.~social security number scams (SSN) with 11 states is much harder than for reward scams with 5 states. We believe that this is a reflection of the optimization of the HMM models: they efficiently extract the available information, with each extracted state supported by a similar level of information. 

We consider these results to successfully demonstrate that effective tools to track scam stages in a live call setting can be built given sufficient data. 
Though strict target accuracy around 50\% may seem low, we note that to achieve this with many states (11 for social security numbers) is substantially difficult as seen in the margin over a random model, and that most of the ``errors'' are in fact identifying a nearby state in the state sequence. 

\begin{table}
    \renewcommand{\arraystretch}{1.2}
    \caption{Accuracy for predicting scam stages and margin over random model.}
    \label{tbl:predict-scam-stage}
    \resizebox{\columnwidth}{!}{%
    \begin{tabular}{l|c|c|c|c|c}
        &  \textbf{Stages} & \textbf{Relaxed} &  \textbf{Margin vs. }&  \textbf{Strict} &  \textbf{Margin vs.}   \\ 
        &         & \textbf{Target}  &  \textbf{Random} &  \textbf{Target} &      \textbf{Random}    \\ \hline
SSN     &      11 &    0.81 &    +0.53 &    0.54 &    +0.45 \\
Refund  &       9 &    0.83 &    +0.49 &    0.59 &    +0.48 \\
Support &       7 &    0.70 &    +0.27 &    0.47 &    +0.33 \\
Reward  &       5 &    0.91 &    +0.31 &    0.50 &    +0.30 \\    
\end{tabular}}
\end{table}

\section{Ethics}

As part of this study, we took the necessary steps to ensure that our research is within legal and ethical boundaries. We consulted with the University's legal experts and Ethics Board to ensure that we were compliant with ethical policy and relevant 
laws. Since scammers are human subjects, we ensured that our actions do not cause them harm. 

It should be noted that the videos that were used in the study were likely made public without the scammers’ consent, however the scammers are completely anonymous, and the content can not be linked to them as individuals. 
Our Ethics Board noted that no requirement to obtain consent was necessary based on the following conditions of our experiment:

\begin{itemize}[noitemsep, topsep=1pt]
    \item Only call recordings already in the public domain were analysed. 
    \item There was no recruitment, targeting or identification of individuals. 
\end{itemize}



\section{Limitations}
\label{sec:limitations}

The sample used for analysis has a number of limitations. Firstly, it only includes YouTube videos of scam calls, most which were recorded by ‘scam-baiters’, people who aim to waste scammers’ time with the full knowledge of the fact that the call was a scam. 
Therefore, the victim is not behaving genuinely in these calls, albeit the scammer is genuine up to the point they realise the victim is a scam baiter. 
In particular, scam baiters sometimes seek to lead the scammer into extended conversations irrelevant to the scam --- they will play the part of an ideal victim in initial phases of a call, but lead the call astray more and more as the call progresses, leading to noisy results towards the later stages of calls. 
Another limitation is that published scam baiting videos are edited during the post-production process. Edits such as merging a few calls in a single video or removing pieces of conversation make the extraction of the scam calls more challenging. Particularly some scam baiters include their own comments (directed to the audience) that pollute the scam call conversation. 
Although efforts have been made to remove these issues by manually removing less relevant parts of the recordings that are not indicative of a typical scam call, as well as splitting scripts containing excerpts from multiple calls, some of these features may persist in the data. 
We also note that some scam baiters may cherry-pick the calls the choose to post on YouTube, however this is probably a good thing, as it acts as a filter for shorter conversations where the scammer quickly recognises that the call is not bona-fide. The inherent bias that results must, however, be acknowledged. 

The capabilities of the video transcribing and diarisation service are also limited and transcripts are prone to errors such as mis-attributing utterances to speakers, incorrectly identifying words used and utterance boundaries, and not recognising very short utterances (instead merging them with the preceding and following utterances). 
In some cases, the diarisation process returned more than 2 speakers which increased the overall workload of correctly identifying scammer's utterances. Again, although efforts have been made to correct these errors during the manual transcript cleanup, some errors still persist. We note that the capabilities of state of the art text to speech services are steadily improving, and expect these issues to be reduced in future. 

One feature of many of the calls that challenges audio transcription is the scammer and scam baiter talking over each other. In these cases, the transcription tool usually identifies just one speaker, occasionally with a word or two injected from the other. This can be extremely difficult to accurately transcribe, even for human transcriptions. 
Sourcing call data where the scammer and victim are in separate audio channels would remove this problem and allow for analysis of the causes and impact of this interesting phenomenon. 

\addition{The data covers relatively few scam types, all of which are from the United States. None the less, it suffices to demonstrate the practicality of recognizing and extracting scam scripts and distinguishing scam types across a selection of common scam types. }
The uneven proportion of the types of scam transcripts in our data set influences the granularity of the information from our analysis. This is particularly visible for reward scams, which had only 23 transcript samples. Although instructive of the need for more data to obtain quality analyses, it remains a limitation of this work.



The fact that the emotion detection model used is a text only model presents another limitation. Such models are not capable of detecting prosodic emotion signals (characteristics related to the tone, pitch, accent and other paralinguistic voice features) and rely solely on the words used. 

\section{Future work}
Our data collection approach enabled us to obtain sufficient data to perform meaningful machine learning pattern recognition on several types of scam. This motivates future approaches to obtain large data sets of scam call conversations. 
Continuing to monitor public releases of scam baiter conversations will no doubt allow further similar conclusions and insights to be gained about newer types of scam. Directly engaging with scam baiters and recruiting people to engage in scam baiting would allow even larger and more timely data sets.

Another approach to obtain greater volumes of data as well as more timely data (for example data for new scam types shortly after they appear) could be to train conversational AI to perform scam baiting via telephony honeypots. 
We believe that recent advances in conversational AI have reached the stage where this has become an effective strategy. 
Criminal organisations that specialise in phone scams are constantly evolving by implementing new technical and social engineering techniques. Moreover, they adjust the scam campaigns to take advantage of recent events (e.g., the pandemic crisis). Unfortunately, law enforcement and even scam-fighting organisations struggle to quickly provide information about the latest scams. We believe that our analysis pipeline could be adjusted to help in identifying and more importantly understanding new scam campaigns, providing valuable insights for both detection and public education campaigns. 

An interesting direction for future investigation is the application of our analysis results into real time detection models and tools. For example, an ongoing conversation that has a high probability score in a certain HMM scam model could raise an alarm. Such technology could be integrated into existing scam prevention applications and promises to provide real-time alerts warning the victim of their immanent mistake.




\section{Conclusion}
Phone scams still remain a popular and inexpensive method to execute scam campaigns. Modern prevention solutions do not address all aspects of scam calls, and thus they are not effective. In this research we analysed 90 hours of scam transcripts acquired from public sources. 
We applied machine learning techniques to identify topics, emotions and patterns in the dynamics of scam calls. 
In particular, we found that Hidden Markov Models over topic model output were effective at identifying steps in underlying scam scripts and techniques 
We further identified aspects of the expression of emotion by scammers, including a general tendency to express little emotion in the scam types in our data, the use of anger to bully the victim into compliance and the use of familiar and emotional chit chat to lull the victim in the last stages before extracting payment from them. 
Finally, we mapped our findings to the literature on social engineering and persuasion techniques and provided an extensive set of examples of their application from our data. 

Our work demonstrates the effectiveness of a combination of structured subjective analysis and automated machine learning techniques to analyse and understand the content and structure of phone scam conversations. We hope that our contribution will enable more effective analysis of scam attempts in the future and contribute to our growing understanding of the phone scam phenomenon and how to defeat it. 


\section*{Acknowledgements}
Research supported by the Australian Office of National Intelligence through National Intelligence and Security Discovery Research Grants (NISDRG) number NI220100105 and the Macquarie University Cyber Security Hub.



\bibliographystyle{IEEEtranS}
\bibliography{zotero-phone-scams-urls,zotero-refs,Remote,ians-refs}

\clearpage

\renewcommand{\thesection}{\Alph{section}}
\section*{APPENDIX}
\setcounter{section}{0}

\section{NLP Model Details for Predicting Scam Type and Scam Stages}
\label{appendix:nlp-model-details}
For both these tasks, we use a standard pre-trained text classifier built on a RoBERTa base model~\cite{liu_roberta_2019} from the Huggingface model repository\footnote{\url{https://huggingface.co/roberta-base}}, which was then fine-tuned on our scam transcript data. We used a learning rate of $2^{-5}$, batch size of 16 (scam state prediction) and 10 (scam type prediction), and weight decay of 0.01. No fine tuning was done and performance was measured via cross validation (average of per-fold accuracy for scam state prediction and per-fold F1 for scam type prediction, see below for details). We used an early stopping strategy with a patience of 5 epochs and threshold of 0.01 on both metrics. 

Data for scam type prediction consisted of scammer utterances from a given scam transcript delimited by RoBERTa's separator token (``</s>''). Data labels were annotated scam types. The number of utterances was progressively increased with a new model trained each time. Separate binary classifiers were trained for each scam type. Cross validation was performed with 7 folds (4 for reward scams) stratified over scam types, F1 used as the evaluation metric and average performance over folds reported. 

Data for scam stage prediction consisted of scammer utterances from a given scam transcript delimited with the string ``<UTT\_\emph{category}\_\emph{time-stamp}>'' followed by RoBERTa's separator token (``</s>''). Here ``\emph{category}'' is replaced by the name of the annotated scam category and ``\emph{time-stamp}'' by the time stamp in the YouTube video corresponding to the start of the utterance. 
Data labels were state labels from the HMM model for the respective scam category (see Section~\ref{sec:HMM}). 
Separate classifiers were trained for each scam category and data consisted only of transcripts from that category. 
We report two forms of accuracy: the first (used for early stopping) accepts the previous and following state as correct, the second only accepts the state assigned to the utterance in question. Note that there are typically long sequences of the same state, so the utterance with the following state may be several utterances ahead. 
Cross validation was performed with 6 folds, chosen such that all data from a given transcript appears in only one fold, and average performance over folds reported.


\vspace{\fill}\newpage
\section{Heat maps of emotion associations with HMM States}
\label{appendix:emo-state-heat-maps}

{
\begin{figure}[ht!]
     \centering
     \begin{subfigure}[b]{0.325\textwidth}
         \includegraphics[width=\textwidth]{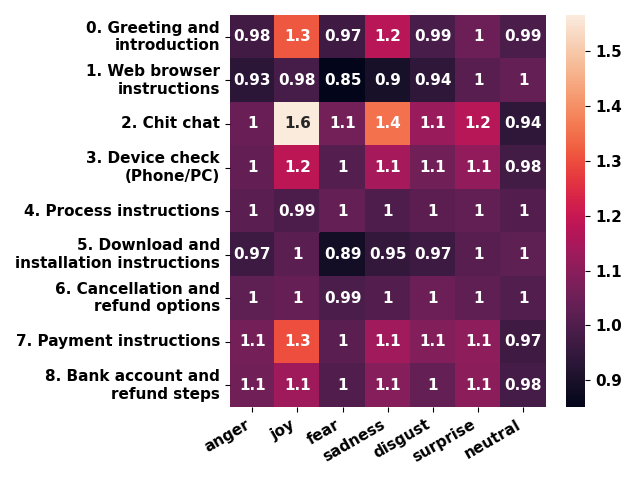}
         \caption{Refund scam emotions heatmap}
         \label{fig:y equals x}
     \end{subfigure}
\hfill
     \begin{subfigure}[b]{0.325\textwidth}
         \includegraphics[width=\textwidth]{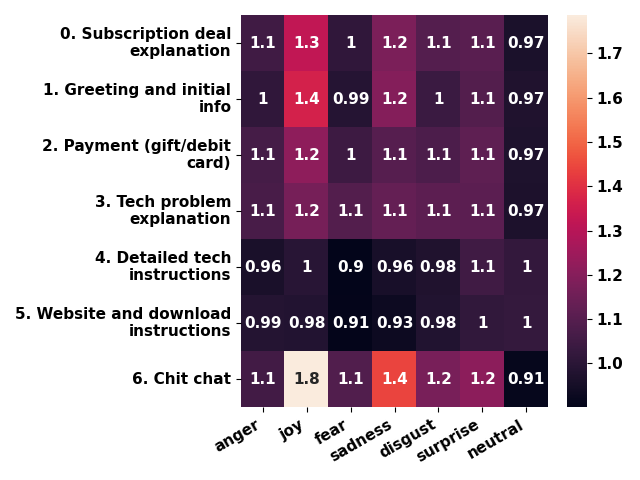}
         \caption{Support scam emotions heatmap}
         \label{fig:three sin x}
     \end{subfigure}
\hfill
     \begin{subfigure}[b]{0.325\textwidth}
         \includegraphics[width=\textwidth]{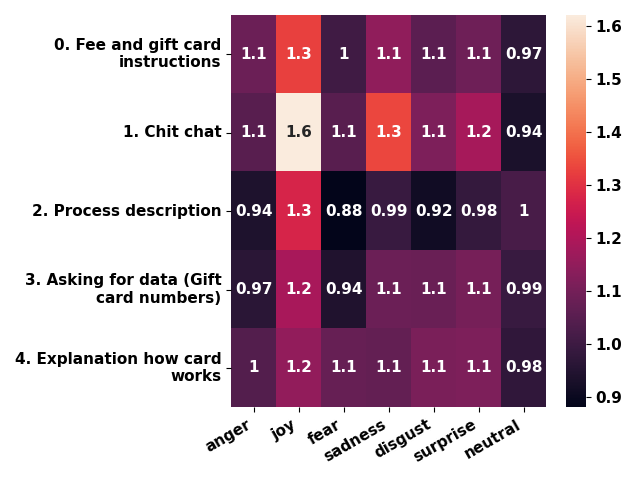}
         \caption{Reward scam emotions heatmap}
         \label{fig:five over x}
     \end{subfigure}
        \caption{Scam types emotion heatmaps, scores are mean for that state relative to median over all utterances.}
        \label{fig:three graphs}
\end{figure}
}

\clearpage
{\onecolumn
\section{Topic descriptions and frequencies}\label{appendix:topic-desc}
We manually labelled utterances with high corresponding topic scores. For each topic, we collected the top 100 utterances. 
We manually inspected the top 10 utterances and 5 utterances randomly selected from the top 100 set (weighted by utterances' topic scores). 
These 15 utterances were used by two of the authors to assign a short description for each topic. 
In the table below, topic frequencies 50\% above a random baseline (i.e.,~$>0.03$, blue) are highlighted, as are topics with merged meanings (pink) and incoherent topics (green). 
\begin{table}[bh]
\footnotesize
\label{tab:topic_description}
\resizebox{.995\textwidth}{!}{%
\setlength{\tabcolsep}{2pt}
\begin{tabular}{c|>{\setlength{\baselineskip}{0.7\baselineskip}}p{0.8\textwidth}|c|c|c|c}

\multirow{2}{*}{\textbf{ID}} &
\multirow{2}{*}{\textbf{Topic description}} &
\multicolumn{4}{c}{\textbf{Frequency}} \\ \cline{3-6}
& & \textbf{Refund} &
\textbf{Reward} &
\textbf{SSN} &
\textbf{Support} 
\\ \hline


0 & names and asking about them & 0.015 & 0.018 & 0.024 & 0.017 \\ \hline
1 & hello, talk to manager, want to talk to, transferred call, can't hear you & 0.019 & 0.017 & 0.021 & 0.020 \\ \hline
2 & monetary amount and payment, including when paid & 0.024 & 0.024 & 0.019 & 0.019 \\ \hline
3 & obtaining gift cards from a nearby store & 0.024 & \cellcolor{blue!25} \textbf{0.047} & 0.016 & 0.023 \\ \hline
4 & free/great deals, cable tv (Roku) subscription, discount vouchers & 0.018 & \cellcolor{blue!25} \textbf{0.039} & 0.013 & 0.024 \\ \hline
\cellcolor{red!15} 5* & sob story asking for money, 'like' as affirmation, about how you like jobs, calling tomorrow, ... & 0.023 & 0.022 & 0.016 & 0.024 \\ \hline 
\cellcolor{red!15} 6* & giving/receiving rebates, rewards ...; 'today', sending information, mailing address; bank login under remote desktop, ... & 0.018 & \cellcolor{blue!25} \textbf{0.054} & 0.015 & 0.016 \\ \hline 
7 & annoyed communication break, 'are you listening', 'angry', 'yelling', 'not understanding' & 0.017 & 0.018 & 0.025 & 0.020 \\ \hline
8 & computer/phone, browser, mobile phone instructions & 0.023 & 0.013 & 0.012 & \cellcolor{blue!25} \textbf{0.030} \\ \hline
9 & short utterances with random words? 'moment', 'okay', ... & 0.021 & 0.018 & 0.020 & 0.022 \\ \hline
10 & cancelling an (amazon/PayPal) order/transaction; accounts, phones and installing applications & 0.023 & 0.015 & 0.012 & 0.023 \\ \hline
\cellcolor{red!15} 11* & formal instructions; instruct to fill a form; 'get some paper'; 'listen to me'; 'go ahead' ..with case, ..and fill; 'let me know' & 0.015 & 0.016 & 0.022 & 0.015 \\ \hline 
12 & stopped services/program/working; computer problems, scans and updates & 0.026 & 0.012 & 0.011 & 0.024 \\ \hline
\cellcolor{red!15} 13* & url instructions; spelling out 'Nancy,delta,mary,charlie; 'write down...' and 'ok?' & 0.023 & 0.014 & 0.016 & 0.021 \\ \hline 
14 & verifies/checks/about(hacked) (bank/amazon) account; describing onscreen context ('see'); email for cancelling account; 'I will be securing your computer'; affirmations 'like/see/okay/right' & 0.024 & 0.014 & 0.015 & 0.020 \\ \hline
15 & banks and bank accounts, 'balance/check account/savings' & 0.020 & 0.014 & 0.026 & 0.018 \\ \hline
16 & clicking instructions, download, free, options & 0.029 & 0.012 & 0.012 & 0.027 \\ \hline
\cellcolor{red!15} 17* & call/call back/hold/stay on the line; drive somewhere; take time out, give time; affirmations (okay, all right); thanks & 0.023 & 0.020 & 0.017 & 0.019 \\ \hline 
18 & (very) familiar chit chat and pleasantries, apologies, familiar farewell, 'honey' as familiar address, happy birthday, ... & 0.024 & 0.027 & 0.018 & 0.025 \\ \hline
19 & describing investigation, 'prove guilty', formal tone, 'listen carefully', 'don't interrupt', ... & 0.012 & 0.012 & 0.028 & 0.012 \\ \hline
20 & Refund/support scam introduction, 'free download' & 0.026 & 0.018 & 0.014 & 0.018 \\ \hline
21 & instructing scammer to interact with computer, asking if victim is in front of computer, cancellation/refund from & 0.023 & 0.014 & 0.016 & 0.018 \\ \hline
22 & details computer instructions, sending/entering a code/password/credit/debit card no., 'cash app', 'please check email' & 0.020 & 0.013 & 0.011 & \cellcolor{blue!25} \textbf{0.031} \\ \hline
23 & DEA/official CASE, 'officer', line recorded, 'evidence', 'better way', 'arrest warrant'... & 0.013 & 0.012 & 0.029 & 0.014 \\ \hline
24 & computer instructions (team viewer), 'do you see the ...?', 'open the..' & 0.029 & 0.011 & 0.011 & 0.025 \\ \hline
25 & instructing to take notes, scammer name and official or case id number; verify victim identity with DOB/SSN & 0.014 & 0.015 & \cellcolor{blue!25} \textbf{0.031} & 0.013 \\ \hline
26 & identity theft scenario, ask for financial information 'to tell which accounts are fraudulent', 'freeze fraudulent accounts' & 0.013 & 0.014 & 0.028 & 0.014 \\ \hline
27 & instructions to enter a url, spelling out letter by letter & 0.027 & 0.013 & 0.013 & 0.025 \\ \hline
\cellcolor{red!15} 28* & describing money laundering / drug trafficking case against victim, list of banks & 0.012 & 0.014 & 0.033 & 0.014 \\ \hline 
29 & payment instructions, wire money, 'do you have a car', drive to the bank/store, what to 'tell' the bank ... & 0.026 & 0.027 & 0.015 & 0.018 \\ \hline
\cellcolor{red!15} 30* & payment instructions, 'get a vanilla' card, 'going to', 'get'; 'your machine is hacked...' & 0.023 & 0.026 & 0.014 & 0.026 \\ \hline 
31 & setting up support scam; 'you've been hacked, you need my help to secure your computer', 'network', 'firewall', 'security' & 0.025 & 0.015 & 0.012 & 0.029 \\ \hline
32 & setting up social security scam; legal enforcement action against your ssn, ... & 0.012 & 0.011 & \cellcolor{blue!25} \textbf{0.033} & 0.012 \\ \hline
33 & spelling names; first and last name, 'can you spell...' & 0.015 & 0.022 & 0.027 & 0.018 \\ \hline
34 & numbers; 'number' & 0.017 & 0.025 & 0.024 & 0.018 \\ \hline
35 & refund scam, filling in a form with PII (team viewer etc.. installed); 'accounts', 'money', 'mom' & 0.027 & 0.015 & 0.011 & 0.016 \\ \hline
36 & introductions, 'thank you for calling', 'how can/may I help you?', 'thank you for holding the line' & 0.019 & 0.019 & 0.024 & 0.020 \\ \hline
37 & short personal utterances, 'I/you know/think', 'people' & 0.023 & 0.027 & 0.021 & 0.028 \\ \hline
38 & about small payments/fee, 'one-time', 'credit/visa card questions', '95 cents', 'send pictures of gift cards' & 0.017 & \cellcolor{blue!25} \textbf{0.063} & 0.016 & 0.021 \\ \hline
39 & introducing ssn scam: 'read out legal charges against you', 'suspend your ssn', 'federally recorded and monitored', 'don't interrupt', 'I have / give you the information' & 0.012 & 0.011 & \cellcolor{blue!25} \textbf{0.033} & 0.013 \\ \hline
\cellcolor{green!15}40* & incoherent topic, short obscure utterances & 0.020 & 0.019 & 0.022 & 0.021 \\ \hline
\cellcolor{red!15} 41* & scam payment instructions: safeguard your money, block the identity thieves by getting amazon cards, stopp your accounts from freezing by buying gov bonds; amazon refund scam explaining final scam step & 0.023 & 0.017 & 0.017 & 0.017 \\ \hline  
\cellcolor{red!15} 42* & don't disclose to/discuss with anyone (before buying gift card/entering bank); stop talking, try to understand, let me speak; calm down; you need to understand; talk to you ... my work; siblings/family members? talking? & 0.017 & 0.019 & 0.024 & 0.019 \\ \hline  
43 & about a car in Texas in your name (with drugs, blood ...), Texas addresses in your name (sometimes linked to the car)... & 0.014 & 0.017 & \cellcolor{blue!25} \textbf{0.034} & 0.014 \\ \hline
44 & have you lost...? (about identity theft), do you know anyone who visited (Texas) recently/who can do this (id theft..)? & 0.014 & 0.012 & \cellcolor{blue!25} \textbf{0.034} & 0.016 \\ \hline
45 & I told/am asking you; tell me (pwd) & 0.019 & 0.025 & 0.022 & 0.022 \\ \hline
46 & asking for PII to 'verify your identity', did you receive an id/case number? & 0.014 & 0.022 & \cellcolor{blue!25} \textbf{0.030} & 0.015 \\ \hline
47 & specific times/durations (I'm going to call, of your time, current time, bank open time...) & 0.024 & 0.023 & 0.015 & 0.020 \\ \hline
\cellcolor{green!15}48* & incoherent, 'like', 'use', 'right', 'know'  & 0.016 & 0.022 & 0.020 & 0.018 \\ \hline
49 & frustratedly repeated instructions, press windows key/letter/..., click accept & 0.024 & 0.013 & 0.011 & 0.022 \\ \hline
\end{tabular}
}
\end{table}
}
{\onecolumn

\section{HMM Transition Graphs}\label{appendix:transition-graphs}
Below are the state transition graphs for refund, support and reward (Figures~\ref{fig:refund-transition-graph},~\ref{fig:support-transition-graph},~\ref{fig:reward-transition-graph}). To determine the thresholds for including edges in the graph, we first discounted the average self-probability of nodes (probability of a node linking to itself) and included links with probability greater than the average remaining probability: 
$threshold=(1-tr(T)/n) / (n-1)$ 
where $T$ is the state transition matrix, 
$tr(.)$ 
is the matrix trace operation and $n$ is the number of states.

\begin{figure*}[ht!]
    \centering
    \includegraphics[width=\textwidth]{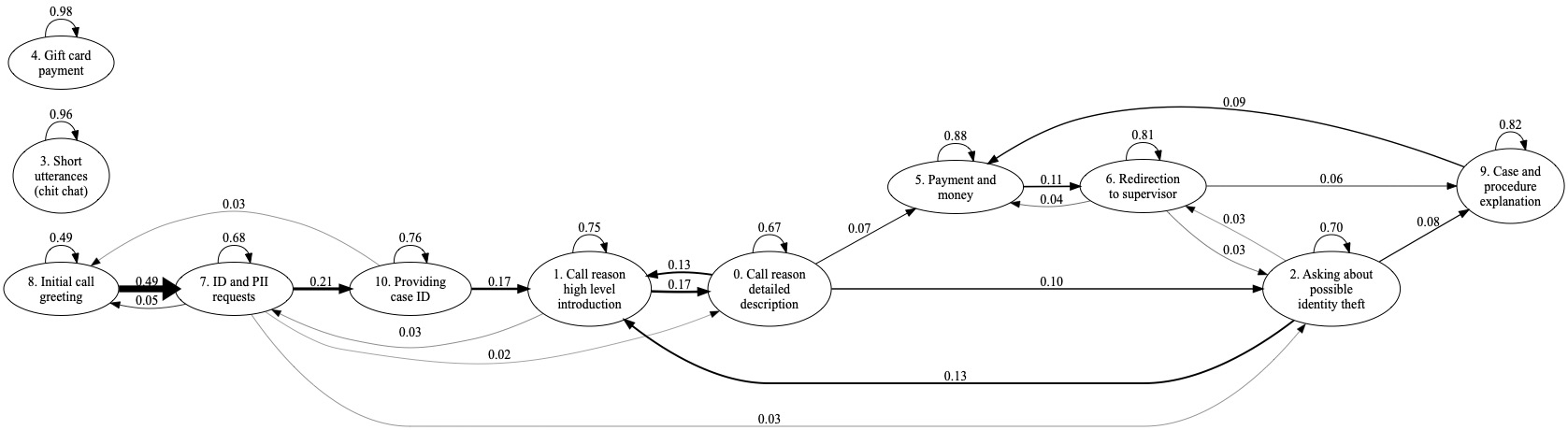}
    \caption{HMM state transition graph for ``social security number'' scams. Numbers on links indicate transition probabilities (those less than 0.023 omitted).}
    \label{fig:ssn-transition-graph}
\end{figure*}

\begin{figure*}[ht!]
    \centering
    \includegraphics[width=\textwidth]{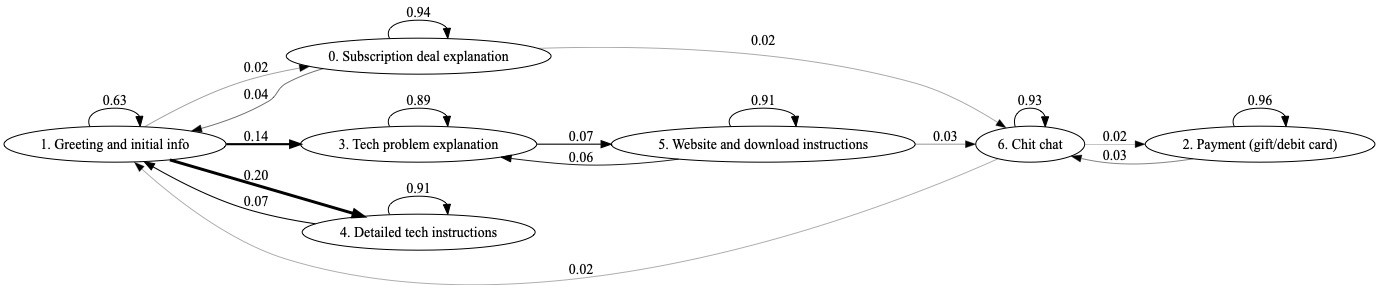}
    \caption{HMM state transition graph for ``support'' scams. Numbers on links indicate transition probabilities (those less than 0.020 omitted).}
    \label{fig:support-transition-graph}
\end{figure*}

\begin{figure*}[ht!]
    \centering
    \includegraphics[width=.4\textwidth]{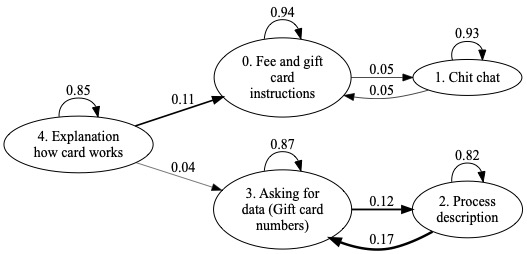}
\caption{HMM state transition graph for ``reward'' scams. Numbers on links indicate transition probabilities (those less than 0.040 omitted).}
    \label{fig:reward-transition-graph}
\end{figure*}
}

{\onecolumn
\section{HMM States interpretations}
\label{appendix:hmm-state-interpretations}

\renewcommand{\arraystretch}{1.2}
\begin{table}[ht!]
\resizebox{\textwidth}{!}{%
\footnotesize
\begin{tabular}{|c p{0.6\textwidth} l|}
\hline
\multicolumn{1}{|c|}{\textbf{State}} & \multicolumn{1}{l|}{\textbf{Top 3 topics}} & \multicolumn{1}{l|}{\textbf{Interpretation}} \\ \hline
\multicolumn{3}{|c|}{\rule{0pt}{12pt}\textbf{Refund scam}} \\[4pt]
\hline
\multicolumn{1}{|c|}{0} & \multicolumn{1}{p{0.7\textwidth}|}{{[}20. 0.234 refund/support intro{]}  {[}36. 0.219 introductions{]}  {[}21. 0.092 computer: in front of, form{]}} & Greeting and introduction \\ \hline
\multicolumn{1}{|c|}{1} & \multicolumn{1}{p{0.7\textwidth}|}{{[}27. 0.300 URL instructions, spelled out (long){]}  {[}24. 0.235 computer: team viewer, do you see?{]}  {[}49. 0.130 computer: frustrated repeated instructions{]}} & Web browser instructions \\ \hline
\multicolumn{1}{|c|}{2} & \multicolumn{1}{p{0.7\textwidth}|}{{[}47. 0.107 times/durations{]}  {[}37. 0.092 short personal utterances{]}  {[}18. 0.086 familiar chit chat{]}} & Chit chat \\ \hline
\multicolumn{1}{|c|}{3} & \multicolumn{1}{p{0.7\textwidth}|}{{[}8. 0.245 browser/phone instructions{]}  {[}10. 0.106 (install app to) cancel order{]}  {[}21. 0.084 computer: in front of, form{]}} & Device check (Phone/PC) \\ \hline
\multicolumn{1}{|c|}{4} & \multicolumn{1}{p{0.7\textwidth}|}{{[}41. 0.130 payment: safeguard money, bonds, gift cards{]}  {[}31. 0.094 support intro: hacked, secure your computer{]}  {[}10. 0.087 (install app to) cancel order{]}} & Process instructions \\ \hline
\multicolumn{1}{|c|}{5} & \multicolumn{1}{p{0.7\textwidth}|}{{[}16. 0.405 instructions to click{]}  {[}24. 0.147 computer: team viewer, do you see?{]}  {[}34. 0.053 numbers{]}} & Download and installation instructions \\ \hline
\multicolumn{1}{|c|}{6} & \multicolumn{1}{p{0.7\textwidth}|}{{[}12. 0.081 stopped services, computer problems{]}  {[}27. 0.070 URL instructions, spelled out (long){]}  {[}9. 0.056 short, random utterance{]}} & Cancellation and refund options \\ \hline
\multicolumn{1}{|c|}{7} & \multicolumn{1}{p{0.7\textwidth}|}{{[}29. 0.124 payment instructions (drive to, wire){]}  {[}3. 0.121 obtain gift card{]}  {[}2. 0.076 \$amount, payment{]}} & Payment instructions \\ \hline
\multicolumn{1}{|c|}{8} & \multicolumn{1}{p{0.7\textwidth}|}{{[}15. 0.122 banks and bank accounts{]}  {[}35. 0.111 fill form with PII{]}  {[}2. 0.088 \$amount, payment{]}} & Bank account and refund steps \\ \hline 
\multicolumn{3}{|c|}{\rule{0pt}{12pt}\textbf{Social security number (SSN) scam}} \\[4pt]
\hline
\multicolumn{1}{|c|}{0} & \multicolumn{1}{p{0.7\textwidth}|}{{[}43. 0.358 car in Texas, blood, drugs{]}  {[}28. 0.246 describing case details, list of banks{]}  {[}19. 0.098 ssn intro: investigations, prove guilty{]}} & Call reason detailed description \\ \hline
\multicolumn{1}{|c|}{1} & \multicolumn{1}{p{0.7\textwidth}|}{{[}39. 0.155 ssn intro: charges, monitored, no interrupt{]}  {[}7. 0.115 anger/annoyance{]}  {[}11. 0.095 fill form, go get paper, case{]}} & Call reason high level introduction \\ \hline
\multicolumn{1}{|c|}{2} & \multicolumn{1}{p{0.7\textwidth}|}{{[}44. 0.496 have you visited Texas/had identity theft?{]}  {[}26. 0.112 ssn: case and procedure explanation{]}  {[}45. 0.058 emphatic 'I tell/told/ask you'{]}} & Asking about possible identity theft \\ \hline
\multicolumn{1}{|c|}{3} & \multicolumn{1}{p{0.7\textwidth}|}{{[}37. 0.171 short personal utterances{]}  {[}5. 0.073 familiar language{]}  {[}42. 0.068 don't disclose/discuss with anyone{]}} & Short utterances (chit chat) \\ \hline
\multicolumn{1}{|c|}{4} & \multicolumn{1}{p{0.7\textwidth}|}{{[}3. 0.139 obtain gift card{]}  {[}41. 0.066 payment: safeguard money, bonds, g. cards{]}  {[}38. 0.059 about small payment/fee{]}} & Gift card payment \\ \hline
\multicolumn{1}{|c|}{5} & \multicolumn{1}{p{0.7\textwidth}|}{{[}15. 0.408 banks and bank accounts{]}  {[}2. 0.117 \$amount, payment{]}  {[}38. 0.072 about small payment/fee{]}} & Payment and money \\ \hline
\multicolumn{1}{|c|}{6} & \multicolumn{1}{p{0.7\textwidth}|}{{[}23. 0.229 ssn: Recorded line or transfer to supervisor{]}  {[}36. 0.103 introductions{]}  {[}1. 0.089 hello, transfer{]}} & Redirection to supervisor \\ \hline
\multicolumn{1}{|c|}{7} & \multicolumn{1}{p{0.7\textwidth}|}{{[}46. 0.295 verify your id with PII{]}  {[}33. 0.161 names: spelling out, first/last{]}  {[}32. 0.107 ssn intro: legal enforcement action{]}} & ID and PII requests \\ \hline
\multicolumn{1}{|c|}{8} & \multicolumn{1}{p{0.7\textwidth}|}{{[}36. 0.334 introductions{]}  {[}0. 0.075 names: asking, spelling out{]}  {[}32. 0.075 ssn intro: legal enforcement action{]}} & Initial call greeting \\ \hline
\multicolumn{1}{|c|}{9} & \multicolumn{1}{p{0.7\textwidth}|}{{[}26. 0.271 ssn: case and procedure explanation{]}  {[}42. 0.149 don't disclose/discuss with anyone{]}  {[}23. 0.090 ssn: Recorded line or transfer to supervisor{]}} & Case and procedure explanation \\ \hline
\multicolumn{1}{|c|}{10} & \multicolumn{1}{p{0.7\textwidth}|}{{[}34. 0.197 numbers{]}  {[}25. 0.162 ssn: write down case id, verify DOB/SSN{]}  {[}32. 0.104 ssn intro: legal enforcement action{]}} & Providing case ID \\ \hline
\multicolumn{3}{|c|}{\rule{0pt}{12pt}\textbf{Support scam}} \\[4pt]
\hline
\multicolumn{1}{|c|}{0} & \multicolumn{1}{p{0.7\textwidth}|}{{[}4. 0.130 deals and discount vouchers{]} {[}9. 0.076 short, random utterance{]} {[}40. 0.070 short obscure utterances{]}} & Subscription deal explanation \\ \hline
\multicolumn{1}{|c|}{1} & \multicolumn{1}{p{0.7\textwidth}|}{{[}36. 0.180 introductions{]} {[}1. 0.157 hello, transfer{]} {[}2. 0.041 \$amount, payment{]}} & Greeting and initial info \\ \hline
\multicolumn{1}{|c|}{2} & \multicolumn{1}{p{0.7\textwidth}|}{{[}3. 0.136 obtain gift card{]} {[}38. 0.085 about small payment/fee{]} {[}2. 0.078 \$amount, payment{]}} & Payment (gift/debit card) \\ \hline
\multicolumn{1}{|c|}{3} & \multicolumn{1}{p{0.7\textwidth}|}{{[}31. 0.186 support intro: hacked, secure your computer{]} {[}12. 0.075 stopped services, computer problems{]} {[}8. 0.071 browser/phone instructions{]}} & Tech problem explanation \\ \hline
\multicolumn{1}{|c|}{4} & \multicolumn{1}{p{0.7\textwidth}|}{{[}22. 0.157 computer: send code/pwd, cash app{]} {[}16. 0.093 instructions to click{]} {[}8. 0.087 browser/phone instructions{]}} & Detailed tech instructions \\ \hline
\multicolumn{1}{|c|}{5} & \multicolumn{1}{p{0.7\textwidth}|}{{[}16. 0.136 instructions to click{]} {[}24. 0.133 computer: team viewer, do you see?{]} {[}27. 0.121 URL instructions, spelled out (long){]}} & Website and download instructions \\ \hline
\multicolumn{1}{|c|}{6} & \multicolumn{1}{p{0.7\textwidth}|}{{[}37. 0.191 short personal utterances{]} {[}18. 0.114 familiar chit chat{]} {[}5. 0.070 familiar language{]}} & Chit chat \\ \hline
\multicolumn{3}{|c|}{\rule{0pt}{12pt}\textbf{Reward scam}} \\[4pt]
\hline
\multicolumn{1}{|c|}{0} & \multicolumn{1}{p{0.7\textwidth}|}{{[}3. 0.234 obtain gift card{]} {[}38. 0.094 about small payment/fee{]} {[}42. 0.076 don't disclose/discuss with anyone{]}} & Fee and gift card instructions \\ \hline
\multicolumn{1}{|c|}{1} & \multicolumn{1}{p{0.7\textwidth}|}{{[}37. 0.141 short personal utterances{]} {[}18. 0.101 familiar chit chat{]} {[}17. 0.066 call back, scammer is serious/generous{]}} & Chit chat \\ \hline
\multicolumn{1}{|c|}{2} & \multicolumn{1}{p{0.7\textwidth}|}{{[}6. 0.311 rebate voucher{]} {[}38. 0.084 about small payment/fee{]} {[}4. 0.079 deals and discount vouchers{]}} & Process description \\ \hline
\multicolumn{1}{|c|}{3} & \multicolumn{1}{p{0.7\textwidth}|}{{[}38. 0.300 about small payment/fee{]} {[}34. 0.092 numbers{]} {[}46. 0.070 verify your id with PII{]}} & Asking for data (Gift card numbers) \\ \hline
\multicolumn{1}{|c|}{4} & \multicolumn{1}{p{0.7\textwidth}|}{{[}30. 0.200 payment instructions (vanilla card){]} {[}4. 0.150 deals and discount vouchers{]} {[}48. 0.107 like, use, right, know{]}} & Explanation how card works \\ \hline
\multicolumn{3}{p{\textwidth}}{\footnotesize Topics are in descending probability ordered. Each topic is enclosed with square brackets. The first number in the brackets is the topic id, followed it's emission probability from the state and it's description. }\\
\end{tabular}
}\end{table}

}
{\onecolumn

\section{Example Transcript with State Transitions}
\label{appendix:hmm-transcript-example}

\renewcommand{\arraystretch}{1.2}
\begin{table*}[!ht]
\footnotesize
    \caption{Example of an social security number scam transcript with HMM states}
    \centering
    \begin{tabular}{|c|c| p{0.85\linewidth}|}
    \hline
        \textbf{State} & \textbf{Timestamp} & \textbf{Utterance} \\ \hline
        8 & 0:01:01 & You reached to the department of social security administration. \textbf{How can I help you?} \\ \hline
        7 & 0:01:18 & ... And the number on which we are talking right now, \textbf{is this your number?} \\ \hline
        7 & 0:01:32 & And did you receive any kind of a \textbf{case ID} or any \textbf{reference ID number} from our department today? ... \\ \hline
        10 & 0:01:43 & No. Okay. Not a problem. So we'll look what I'm saying to you that right now, to give you the information and to give you the \textbf{case ID} number, I have to pull out your case file and I need to verify you then only I'm able to give you the information, which is very important for you. Okay? Okay, great. So first of all, tell me what your \textbf{date of birth}. \\ \hline
        10 & 0:02:12 & Okay. And first I was like Yardi reason, confirm me, which is \textbf{social number} so that I'm able to give you the information. Okay. \\ \hline
        10 & 0:02:38 & And can you tell me that from which state your social was issued to you? Uh, Paris, Texas. Okay. So no. ...  First of all, write down my name. My name is officer Alan Foster. ... Now right on my \textbf{federal employee ID number} means my \textbf{badge ID number} is 5 6 3 2 1 0 1. Okay. Okay. Now write down your \textbf{case ID number} and write it very properly because it's very important for you. Your case ID number is D C 0 3 3 4. ... \\ \hline
        1 & 0:04:31 & Correct. ... So I ordered a request. You, when I will go ahead and \textbf{read out the legal charges against your name}, please do not interrupt me in between. You will be given a fair chance once I'm done with the information. ... \\ \hline
        1 & 0:05:03 & ... It's been federally recorded and monitored. So I would request, you will go ahead and read all the \textbf{legal charges against your name} ... So before I start reading the, $<$inaudible$>$ just a quick question to you \textbf{that your name and your information} has been found to species in the state of Texas. So right now, are you in the state of Texas or where you are right now in the \textbf{state of Texas}? Yes. Okay. And you are, you are from which city are you from? \\ \hline
        1 & 0:06:04 & And have you ever been to the \textbf{state of Pennsylvania} recently or in the past? \\ \hline
        0 & 0:06:16 & ... We are taking this issue. We are taking this issue to the FBA headquarters ... The investigation started when we found an abandoned car on the south border. ... \textbf{First of all, let me give you the information that exactly what happened and what's going on}. So we are taking this issue to the FBA headquarter, as we are having a \textbf{strong evidence that are enough to prove you guilty} inside the co-taught. The investigation started when we found an abandoned car on the south border of Texas and the car contained some blood. \\ \hline
        0 & 0:07:19 & So look, the investigation started when we found in a bottle on the south border of Texas and the car contained some blood and drug residues inside. Ed has the outdoor investigation. We spunk that the \textbf{car was rented on your name and on your personal social information}. ... we have, uh, recovered 20 pounds of cocaine, ... and some documentation from the financial Institute, ... \textbf{And this entire paperwork has your name on it}. ... \\ \hline
        \multicolumn{3}{l}{\footnotesize Only scammer utterances are presented in the table. Transcript extracted from video with Vohzz6Fgt8M id.}\\
    \end{tabular}
\end{table*}

}
\end{document}